\documentclass[aip,graphicx]{revtex4-1}
\usepackage{graphicx}

\usepackage{hyperref}
\usepackage{amsmath}

\usepackage{color}
\usepackage{bm}
\usepackage[normalem]{ulem}

\newcommand{\bfm}[1]{\textbf{#1}}

\newcommand{\R}[1]{{\color{black}#1}}

\newcommand{\Ohmmm}{$\Omega$m$^2$ }

\newcommand{\half}{\frac{1}{2}}
\newcommand{\dif}{{\rm d}}
\newcommand{\dvol}{{\rm d}^3\bfm{r}}

\newcommand{\vJ}{\bfm{J}}

\newcommand{\vE}{\bfm{E}}
\newcommand{\vB}{\bfm{B}}
\newcommand{\vA}{\bfm{A}}

\newcommand{\vr}{\bfm{r}}

\newcommand{\ten}[1]{\overline{\overline{#1}}}

\newcommand{\vn}{\bfm{n}}

\begin{document}

\title{Screening currents increase thermal quench propagation speed in ultra-high-field REBCO magnets} 

\author{Enric Pardo}
\email[]{enric.pardo@savba.sk}
\affiliation{Institute of Electrical Engineering, Slovak Academy of Sciences, Bratislava, Slovakia}

\author{Anang Dadhich}
\affiliation{Institute of Electrical Engineering, Slovak Academy of Sciences, Bratislava, Slovakia}

\author{Nikola Jerance}
\affiliation{Universit\'e Paris-Saclay, CEA, 91191, Gif-sur-Yvette, France}

\author{Philippe Fazilleau}
\affiliation{Universit\'e Paris-Saclay, CEA, 91191, Gif-sur-Yvette, France}

\date{\today}

\begin{abstract}
Superconducting REBCO ($RE$Ba$_2$Cu$_3$O$_{7-x}$, where $RE$ is a rare earth, typically Y, Gd or Eu) electromagnets are useful for many applications like medical magnetic resonace imaging (MRI), nuclear magnetic resonance (NMR) spectroscopy, and magnets for particle accelerators and detectors. REBCO magnets are also the core of many nuclear fusion energy start-ups. In order to avoid permanent damage during operation, magnet design needs to take electrothermal quench into account, which is due to unavoidable REBCO tape or magnet imperfections. However, most high-field magnet designs do not take superconducting screening currents into account. In this work, we show that it is essential to consider screening currents in magnet design, since they highly speed-up electrothermal quench propagation. Our study is based on detailed numerical modeling, based on the Minimum Electromagnetic Entropy Production (MEMEP) and Finite Differences (MEMEP-FD). Benchmarking with well-established Partial Element Equivalent Circuit (PEEC) model supports the correctness of MEMEP-FD. This work focusses on a 32 T all-superconducting magnet design and we analyze in detail the time evolution of electrothermal quench. Our findings will have an impact in the design of ultra-high-field magnets for NMR or user facilities, and possibly for other kinds of magnets, like those for fusion energy.
\end{abstract}


\maketitle 

\section{Introduction}

Superconducting electromagnets are able to generate intense static magnetic fields up to 45 T \cite{hahnS2019Nat}, thanks to the high current density that they can transport. High magnetic fields are necessary for several applications like Magnetic Resonance Imaging (MRI) \cite{vedrineP2015IES, parkinsonBJ2017SST, batesS2023MRM}, Nuclear Magnetic Resonance (NMR) spectroscopy \cite{iwasaY2015IES, iguchiS2016SST, wikusP2022SSTa, yanagisawaY2022SST} and user magnets for physics and material research \cite{weijersHW2013IES, awajiS2017SST, hahnS2019Nat, huX2022SST, fazilleauP2020Cry, kimJ2020RSI, liuJ2020SST, yanagisawaY2022SST, pugnatP2022IES, durochatM2024IES}. Other interesting applications are particle accelerator magnets \cite{rossiL2012RAS, vannugteren2018SST, wangX2019Ins, botturaFP2022FiP} or magents for nuclear fusion energy, such as tokamaks and stellators \cite{mitchellN2021SST}. REBCO ($RE$Ba$_2$Cu$_3$O$_{7-x}$, where $RE$ is a rare earth, typically Y, Gd or Eu) high temperature superconductors (HTS) are the most interesting materials for the highest magnetic field range of these applications, thanks to the high current capacity at high magnetic fields. Indeed, fusion energy start-ups are consuming thousands of kilometers of REBCO tape per year.

All superconducting magnets are prone to electo-thermal quench during operation, such as those caused by tape segments with reduced critical current, $I_c$. If the current overcomes the local $I_c$, heat is generated by Joule effect, which increases temperature, which reduces $I_c$, and further increases heat generation; resulting in a positive feedback loop. Then, magnets need to be designed in order to be reliable against electrothermal quench. Presently, high-field magnet design neglects screening currents for electrothermal quench. This is partly because superconducting screening currents are negligible in classical low-temperature superconducting (LTS) wires like those made of NbTi or NbSn$_3$, which are multi-filamentary \R{and twisted} \cite{wilson, sharmabook}; and partly because of the computational complexity of taking screening currents into account. \R{Although there has been a lot of research on filamented REBCO tapes, filamentation is only effective after transposition, otherwise the wire behaves as a single filament regarding screening currents and AC loss \cite{wilson}. Filamentation in HTS tapes requires cabling, such as in round cables or twisted stacks. However, these cables are not applicable to ultra-high field magnets, which have relatively low bore, because they reduce the engineering current density and they are more sensitive to mechanical forces, among other issues.}

The aim of this work is to show that screening currents highly enhance the speed of electrothermal quench propagation in REBCO high-field magnets. This is due to the relatively high width of the REBCO tapes (4-12 mm) compared to $\sim 20$ $\mu$m of superconducting filaments in LTS. In this article, we analyze by means of numerical modelling the electrothermal quench propagation in a 32 T magnet design with a REBCO insert and we compare the results when considering superconducting screening currents and neglecting them. This magnet is one of the designs for the 32 T and 40 T magnets \cite{durochatM2024IES} in project SuperEMFL \cite{superEMFL}. As quench protection, we consider that the input current source cannot exceed a certain voltage (1 V in our case) \R{and that the current source is connected directly to the insert. Since the voltage in the whole coil is limited, then the current drops when there appears a significant effective resistance due to electrothermal quench. This effective resistance appears when the critical current of any turn in the insert falls below the coil current.} In addition, we consider metal insulation between pancake turns as additional quench protection \cite{fazilleauP2020Cry, lecrevisseT2022SSTa}. Metal-insulated coils work with a similar principle as non-insulated coils \cite{hahnS2012IES, yoonS2016SST} but present higher surface resistance between turns, which increases the voltage during quench. In addition, metal-insulated coils are more reliable against the high stress caused by Lorentz forces.

Although there exist multiple publications about electrothermal quench modelling in REBCO tapes \cite{bonnardCH2017SST, lacroixC2017SST} and coils \cite{stenvallA2023book, liY2014JCS, botturaL1996JCP, badelA2019SST, cavallucciL2021SST, dongF2022APL, chenJ2023SST, vitranoA2023IES, gavrilinAV2021IES, ravaioliE2022IES, genotC2022IES, fazilleauP2024IES, wangYN2024SST, tangY2025HMT}, this is the first work about the impact of screening currents in \R{metal-insulated (or non-insulated)} high-field magnets (preliminary results of this study have been presented at several conferences with slides available at \cite{pardoE2024HTSmod}). However, references \cite{dongF2022APL, tangY2025HMT} take screening currents into account for a single pancake and reference \cite{gavrilinAV2021IES} partially takes screening currents into account but only for the change in critical current due to mechanical bending in a magnet. \R{The most notable existing work is reference \cite{wangYN2024SST}, which analyzes an HTS insert for a high-field magnet. However, modeling in that work assumes that the HTS winding is insulated, while electrothermal quench in metal-insulated coils highly differs from insulated. In addition, reference \cite{wangYN2024SST} does not consider decay current due to voltage limitation. Although computations took screening currents into account, that work does not discuss about the role of screening currents in electrothermal quench when comparing with modeling that assumes uniform current density. This comparison is of high interest for the magnet technology field, since the uniform current density approximation is still common practice in electrothermal quench analysis in magnet design.}

This article is structured as follows. First, we detail the magnet configuration (section \ref{s.conf}). Then, we outline the numerical methods (section \ref{s.method}). The main results are from a method based on the Minimum Eletro Magnetic Entropy Production (MEMEP) \cite{pardoE2016SST, pardoE2017JCP, pardoE2024SSTa} coupled to a Finite Difference (FD) method \cite{dadhichA2024SSTa}, MEMEP-FD. Different from our previous work \cite{dadhichA2024SSTa}, the model in this article is able to analyze quench in high time resolution (down to 10$^{-7}$ s) and take voltage limiting into account. We also improved the computing time. In addition, we use a Partial Element Equivalent Circuit (PEEC) model for benchmarking purposes. We present the results in section \ref{s.results}; which contains a benchmark with the PEEC model for a double pancake, as well as quench analysis of the 32 T magnet. Results support the conclusion that screening currents highly speed-up electrothermal quench (section \ref{s.conclusion}). In the appendix, we present details of the material properties, the MEMEP-FD numerical model, the benchmark between MEMEP-FD and PEEC, and a comparison of the time evolution of electrothermal quench of two different REBCO tapes (Fujikura and Theva). The latter also discusses the details of the electrothermal quench evolution.

\section{Magnet configuration}
\label{s.conf}

\begin{figure}[tbp]
{\includegraphics[trim=0 0 -10 7,clip,height=7 cm]{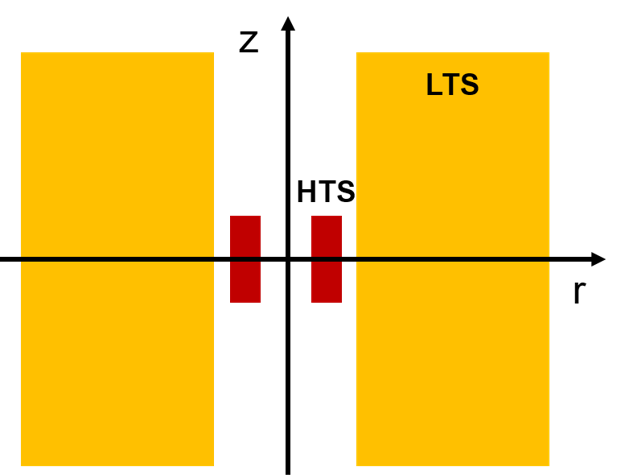}}%
{\includegraphics[trim=280 117 230 137,clip,height=7 cm]{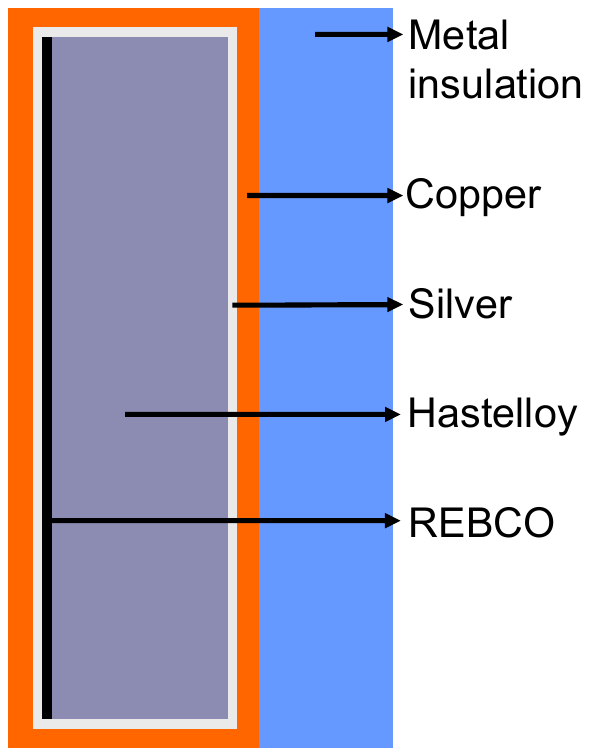}}%
\caption{(Left) Qualitative sketch of the 32 T high-field magnet cross-section. The HTS insert is made of REBCO coated conductor. \R{(Right) Sketch of the HTS tape cross-section with the co-wound metal-insulation layer on the right (not to scale).} \label{f.sketch}}%
\end{figure}

\begin{table}[tbp]
\caption{Main parameters of the REBCO insert and the LTS outset. The calculated critical current, $I_c$, of the insert is 472 A for the Theva APC tape (and 371 for the Fujikura tape).}
\begin{tabular}{ll}
\hline 
\hline 
{\bf HTS insert} & \\
Generated magnetic field & 13 T \\
Rated current & 333 A \\
Inner diameter & 50 mm \\
Outer diameter & 102.5 mm \\
Number of pancakes & 16 \\
Number of turns per pancake & 250 \\
G10 spacer thickness between pancakes & 0.5 mm \\
Contact resistance between turs & $10^{-6}$ $\Omega$m$^2$ \\
\hline
{\bf HTS tape} \\
REBCO superconductor thickness & 1 $\mu$m \\
Hastelloy thickness & 54 $\mu$m \\
Copper and silver stabilization thickness & 20 $\mu$m \\
Durnomag metal insulation & 30 $\mu$m \\
Tape width & 6 mm \\
\hline
{\bf LTS outsert} & \\
Generated magnetic field & 19 T \\
Cold bore diameter & 150 mm \\
Producer & Oxford Instruments \\
\hline
\hline 
\end{tabular}
\label{t.insert}
\end{table}

In this article, we analyze an HTS REBCO insert for a 32 T magnet with the parameters in table \ref{t.insert} (figure \ref{f.sketch}). For the conductor, we consider Theva Advanced Pinning Center (APC) tape. In the appendix, we also present results for Fujikura tape. Both tapes have a Hastelloy substrate, with copper stabilization from all sides and an additional silver layer. Modeling assumes that the electrical and thermal properties of Hastelloy and the metal-insulation material Durnomag are the same as Stainless Steel (SS), since for the latter there is available all necessary experimental data for the required temperature and magnetic induction range and it does not significantly differ from Hastelloy. Similarly, we assume that silver behaves the same as copper (more details in the appendix). Using inputs with refined accuracy will not change the main conclusion of this article.

As quench protection, the magnet uses metal-insulated pancakes at the insert and voltage limitation of 1 V from the current source. Here, we consider a turn-to-turn \R{contact} resistance of 10$^{-6}$ \Ohmmm in the metal insulation, which is of the same order of magnitude of measured \R{values} in metal-insulated coils \cite{genotC2022PhD}. \R{The turn-to-turn resistance, $R_t$, is related to the surface resistance as $R_t=R_s/(2\pi r_iw)$, where $r_i$ is the radial position of the interface between two turns and $w$ is the tape width.}

The current, $I$, of the insert is ramped up to 308.8 A (65.5 \% of the insert $I_c$) at a ramp of 1 A/s. Once the magnet is charged, we assume that a quench event occurs after 162.65 s. \R{In the whole process, we consider voltage limitation of 1 V. This value is above the inductive voltage during ramp up. Then, voltage limitation only plays a role during quench, since the voltage in the coil terminals can be high due to transition to normal state in part of the coil (or regions with angular current well above the critical current), as seen in figure \ref{f.jJc}. As a consequence the net current will decrease (see figure \ref{f.It}).}

\R{
We consider that the HTS insert lies within a 19 T / 150 mm bore Oxford Instruments magnet. For that purpose, we consider the actual geometry of the LTS winding, and hence any field non-homogeneity or radial component is taken into account. Quench of the HTS within the LTS behaves differently than stand-alone. The reasons are the following. First, the magnetic field is larger, which decreases the critical current density. Indeed, the Theva tape is sensitive to both axial and radial components of the magnetic field (see figure \ref{f.JcBT}). Second, the LTS outsert increases the magnetic energy. Since during quench the current decreases to almost zero, most of the magnetic energy converts into heat. Finally, quench in the HTS could induce quench in the LTS, which could influence back the HTS. However, in this article, we assume that the LTS does not quench in order to analyze in detail the HTS behavior.
}

\section{Numerical method}
\label{s.method}

\subsection{{MEMEP-FD}}
\label{s.MEMEP}

The main results of this article are obtained by coupling the electromagnetic model based on the Minimum Minimum Electromagnetic Entropy Production (MEMEP) \cite{pardoE2017JCP, pardoE2016SST, pardoE2024SSTa} with a finite difference method (FD), as done in \cite{dadhichA2024SSTa}. We name the whole coupled method as MEMEP-FD. \R{MEMEMP and FD are coupled using the field-separation iteration method, as detailed in Appendix \ref{s.FD}.}

In more details, the electromagnetic model MEMEP obtains the local angular current density, $J_\varphi$, and radial current density, $J_r$, taking screening currents into account for $J_\varphi$. Here, we further develop the numerical method in \cite{pardoE2024SSTa} to take voltage limitation into account. When the voltage, $V$, overcomes the desired limit, $V_{\rm lim}$; the software modifies the input current $I$ until the difference between $V$ and $V_{\rm lim}$ is below a certain tolerance (1 \% of $V_{\rm lim}$ in our case). In this article, we assume adiabatic conditions. However, the thermal model can also take liquid helium cooling into account \cite{dadhichA2024SSTa}. 

In this work, we assume axial symmetry. This implicitly assumes that when a defect occurs it appears in the whole length of at least one turn. Although this cannot consider defects in a portion that is substantially shorter than one turn, the axi-symmetric model is sufficient for the main conclusion of this work, which is that screening currents highly enhance quench propagation.

In this work, we also improved the thermal part of MEMEP-FD and its coupling. Now, when quench occurs, the time step self-adapts from a starting value of 10$^{-7}$ s to higher time steps, which reach several seconds when the whole magnet is completly quenched. The time step increases when there are less than \R{or equal to} 2 iterations in one time step and it decreases when there are 3 or more iterations\R{, with a minimum time step equal to the initial time step. Self-adaptive time step highly reduces the computing time while keeping the same accuracy}. We checked that calculations for uniform time step of 0.5 ms provide essentially the same result. \R{The optimum performance is for self-adapting time step and initial step of 0.5 ms. For that case, the computing time is of around 7 hours in a desktop computer with CPU AMD Ryzen 9 7950X and 64 MB of RAM memory for a time evolution of up to 2.5 s after damage, while for fixed time step of 0.5 ms the computing time is around 34 h. The features that limit computing speed is the electromagnetic modeling with voltage limitation (computations with no voltage limitation are much faster), and the still relatively short time stepping. The computing time for only electromagnetic modeling from zero-field cool to just before the appearance of the degraded turns is only around 7 minutes.}

\subsection{{Homogenization}}
\label{s.hom}

\begin{figure}[tbp]
{\includegraphics[trim=3 8 9 5,clip,width=12 cm]{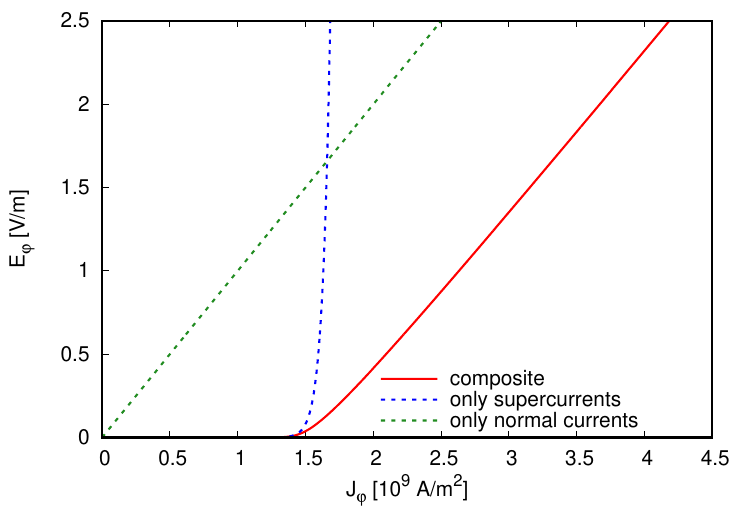}}%
\caption{\R{Qualitative behavior of the assumed homogenized  $E_\varphi-J_\varphi$ relation of (\ref{JE}) for the superconductor-normal composite. The graph also shows the relation assuming if there are only supercurrents or only normal currents (first and second term of (\ref{JE}), respectively). Example for $\rho_n=10^{-9}$ $\Omega$m and $J_{ce}=1.2\cdot 10^{9}$ A/m$^2$.} \label{f.EJ}}%
\end{figure}

In order to reduce computing time, we consider homogenization of all materials composing the HTS coated conductors, which contain a REBCO film and several metal layers \R{(figure \ref{f.sketch}(right))}. For the electric properties, we use the accurate parallel consideration of \cite{pardoE2023IES}. The engineering current density, $J_\varphi$, has a contribution from the \R{normal currents and the superconducting currents (or supercurrents)} as follows
\begin{equation} \label{JE}
J_\varphi=J_{ce}(|E_\varphi|/E_c)^{1/n}{\rm sign}(E_\varphi)+{E_\varphi / \rho_{\rm n}},
\end{equation}
\R{where the first term is for the supercurrent and the second is the normal-current contribution. In the first term,} $J_{ce}$ is the engineering critical current density, $E_c=10^{-4}$ V/m and $n$ is the power-law exponent, which in this work we take as $n=30$. \R{As detailed below, $J_{ce}$ depends on temperature, magnetic field, and its direction. In (\ref{JE})}, we assumed a power-law relation between the $E$ and $J$ \R{of the supercurrents. Our models} can take more sophisticated $E(J)$ \R{relations} of the supercurrents into account, such as those in \cite{rivaN2021SST}, by appropriately modifying the \R{first} contribution in (\ref{JE}). \R{However, there are missing experiments of the $E-J$ relation under both high magnetic fields and high electric fields. Alternatively, the Critical State Model (CSM) corresponds to the limit of $n\to \infty$. For the normal conducting contribution in (\ref{JE}),} $\rho_{\rm n}$ is the homogenized parallel resistivity of all normal conducting materials,
\begin{equation} \label{rhon}
\rho_{\rm n}=\left ( \sum_{m=1}^{N} \frac{S_m}{S\rho_{{\rm n}m}} \right )^{-1},
\end{equation}
where $N$ is the number of different materials, $S_m$ is the cross-section of material $m$, $\rho_{{\rm n}m}$ is its resistivity, and $S$ is the total cross-section \R{(see figure \ref{f.sketch}(right)). Actually, the REBCO layer contains normal-conducting currents in parallel to the supercurrents. This normal contribution is always present, whether the temperature is above the critical temperature, $T_c$, or not. Then, in $\rho_{\rm n}$ of (\ref{rhon}), we also include the contribution from the normal-conducting resistivity of the superconductor, in addition to the other normal-conducting materials. Therefore, in absence of metals the $J_\varphi-E_\varphi$ relation of the superconductor is still (\ref{JE}) but with $\rho_{\rm n}=\rho_{\rm ns}$, being $\rho_{\rm ns}$ the normal-state resistivity of the superconductor. Typically, $\rho_{\rm ns}$ is of the order of 10$^{-7}$ $\Omega$m \cite{royF2010PhD}. Then, when the superconductor becomes normal, relation (\ref{JE}) still holds, but with $J_{ce}=0$. The qualitative $J_\varphi-E_\varphi$ relation is in figure \ref{f.EJ}.} 

For the radial engineering current density, we assume that the effective resistivity is dominated by the contact resistence, and hence
\begin{equation} \label{JrEr}
J_r=E_rd/R_s
\end{equation}
where $R_s$ is the contact resistance between turns and $d$ is the total tape thickness. \R{In this work, we assume $R_s=10^{-6}$ \Ohmmm, which is typical for metal-insulated windings.}

For the thermal properties, we consider that the heat capacity per unit volume, $c_p$, is the average of all materials. The heat conductivity in the axial direction, $k_z$, considers that all layers conduct in parallel, while the radial direction $k_r$ considers that the layers conduct in series. \R{For simplicity, we consider that the tape is made of stacked layers of different materials, and hence we neglect the effect of the stabilization layers on the narrow tape sides (top and bottom of figure \ref{f.sketch}(right)). Then, 
\begin{eqnarray}
&& c_{pe}=\sum_{m=1}^N c_{pm}\frac{d_m}{d} \\
&& k_{ze}=\sum_{m=1}^N k_{m}\frac{d_m}{d} \\
&& k_{re}=\left ( \sum_{m=1}^N \frac{1}{k_m}\frac{d_m}{d} \right )^{-1}\frac{1}{\alpha},
\end{eqnarray}
where $c_{pm}$, $k_m$, $d_m$ are the heat capacity per unit volume, heat conductivity, and thickness of material $m$; $d$ is the total tape thickness; and $N$ is the number of materials. For the heat conductivity in the radial direction, we consider an experimental factor $\alpha=5$ in order to agree with experimental data of heat conductivity of stacks of tapes \cite{durochatM2024IES}. The origin of this factor is the interface heat resistance between layers and between turns. This factor depends on the specific tape and winding method. Although the factor above is not for the same tape producer, it represents the order of magnitude of decrease of $k_{re}$ due to interface thermal resistances.} For all materials, we take the dependence on temperature, $T$, and magnetic induction, $B$, into account; in particular, $J_c(B,
\theta,T)$, $\rho_n(B,T)$, $k(B,T)$ (more details in the appendix). \R{Although we homogenize the properties of the superconducting tape, we consider distict thermal properties of the tape and the G10 spacer between pancakes.}

In order to further improve computing time, we make an additional homogenization among neighboring turns. This assumes that for a given axial position, $z$, $J_\varphi$ and $J_r$ are the same for each group of $n_{\rm nei}$ neighboring turns. In this work, we use $n_{\rm nei}=10$, unless stated otherwise.

\subsection{PEEC model}
\label{s.peec}

In this article, we also consider an additional model for benchmarking purposes. This model employs a coupled approach where the Partial Element Equivalent Circuit (PEEC) method resolves electromagnetic circuits, while finite difference or finite element methods handle thermal analysis. This coupling enables comprehensive simulation of electromagnetic devices where Joule heating from the electromagnetic domain serves as heat sources for thermal calculations.

Originally developed by A. E. Ruehli in the early 1970s \cite{ruehliAE1974IES}, the basic PEEC method transforms electromagnetic problems into equivalent circuit representations by discretizing conducting structures into elementary volumes with partial resistances and inductances, elements derived from Maxwell's integral equations. This approach offers selective discretization advantages, requiring meshing only of conducting materials rather than the entire computational domain.

The PEEC-H (Homogenized) method implements a 2D planar strategy for pancake coils, grouping multiple elements in the radial direction to reduce computational complexity while enabling complete magnet modeling. This homogenization significantly accelerates calculations for large-scale systems.

The PEEC-S (Simplified) method utilizes a 2D axisymmetric formulation where each element represents one or multiple conductor turns. This lumped parameter approach is particularly effective for solenoidal geometries, achieving substantial computational efficiency through dimensional reduction while maintaining essential electromagnetic behavior.

\R{
\subsection{About the normal zone and quenched zone in superconductor-normal composites}
\label{s.Jquench}

Following Wilson \cite{wilson}, the ``normal zone'' in tapes, wires or cables with superconductors and normal conductors in parallel for a certain engineering current density $J$ is defined as the region where the heat generation, $p$, is above half the heat generation in normal state, $p_n$. Then,
\begin{equation} \label{pquench}
p \ge \frac{1}{2}p_n.
\end{equation}
where $p_n=\rho_nJ^2$. The reasoning of the criterion above is that for $p>p_n/2$, the behavior of the heat generation of the superconductor-normal composite is closer to when all current is normal ($J_c=0$, and hence $p=p_n$) than when all current is superconducting ($J_c\gg |J_\varphi|$, and hence $p\ll p_n$). Wilson assumed the Critical State Model (which can be thought as a power-law $E-J$ relation with infinite power-law exponent) and $J_c(T)$ dependence of $J_c(T)\approx J_{c0}(T_c-T)/(T_c-T_0)$ to obtain a sharing temperature, $T_s$, where above it $p\ge p_n/2$. Then, the ``normal zone'' was defined as the region with $T\ge T_s$. This sharing temperature is below the critical temperature, $T_s<T_c$. Strictly speaking, since $T_s<T_c$ the superconductor is still in superconducting state, in the sense that the material still presents superconducting currents, whose macroscpic electrical properties are dominated by vortex pinning and vortex dynamics. Indeed, $J_c$ does not vanish at $T=T_s$. In this work, we refer to the aforementioned ``normal zone'' as the ``quenched zone'' in order to avoid confusion.

Next, we generalize the definition of the ``quenched zone'' for any $J_c(T)$ relation. Actually, (\ref{pquench}) already defines the quenched zone universally. In order to simplify the interpretation, we will consider the angular currents only; since radial currents also cause heat generation. Then, the ``quenched zone'' is defined as
\begin{equation} \label{pphiquench}
p_\varphi\ge \frac{1}{2}p_{\varphi n},
\end{equation}
where $p_\varphi$ is the generated heat due to $J_\varphi$ and $p_{\varphi n}$ is its value in normal state, $p_{\varphi n}=\rho_nJ_\varphi^2$. For the qualitative interpretation of the results, we can assume the Critical State Model (CSM). For the CSM, the $E(J_\varphi)$ relation (\ref{JE}) above becomes (limit $n\to \infty$)
\begin{equation} \label{EJ_CSM}
E_\varphi\approx \begin{cases} 
					\rho_n(|J_\varphi|-J_{ce}){\rm sign}(J_\varphi) & \text{if $|J_\varphi|\ge J_{ce}$}\\
					0 & \text{if $|J_\varphi|< J_{ce}$}
					\end{cases} .
\end{equation}
Using that $p_\varphi=J_\varphi E_\varphi$ and $E_\varphi$ from (\ref{EJ_CSM}), we obtain that the quenched zone criterion of (\ref{pphiquench}) is equivalent to 
$\rho_n |J_\varphi| (|J_\varphi|-J_{ce}) \ge \rho_n|J_\varphi|^2/2$, and hence
\begin{equation} \label{Jquench}
|J_{\varphi}|\ge 2J_{ce}.
\end{equation}
Then, this condition is equivalent to Wilson's $T\ge T_s$ for the quenched zone criterion of (\ref{pphiquench}).

In this article, we compute $J_\varphi$, $J_r$ and $T$ following the numerical model in sections \ref{s.MEMEP}-\ref{s.peec} for the $J_\varphi - E_\varphi$ relation in (\ref{JE}). We apply the condition (\ref{Jquench}) only for a qualitative analysis of the ``quenched zone''.
}

\section{Results}
\label{s.results}

\begin{figure}[tbp]
{\includegraphics[trim=0 0 0 0,clip,width=8.2 cm]{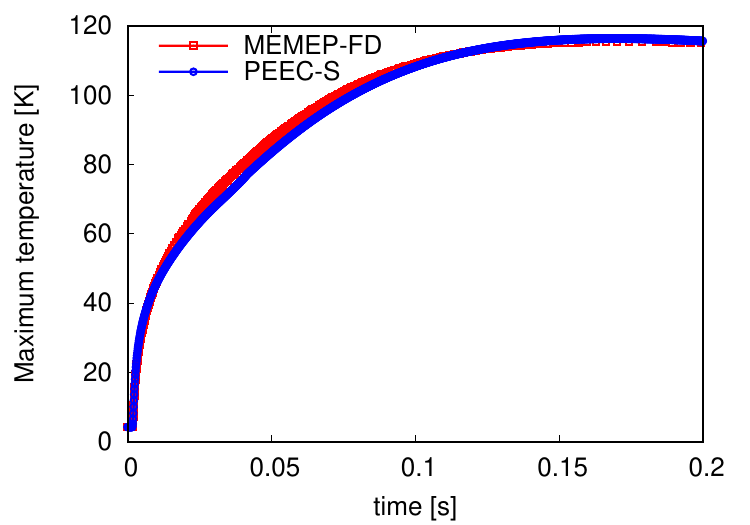}}%
{\includegraphics[trim=0 0 0 0,clip,width=8.2 cm]{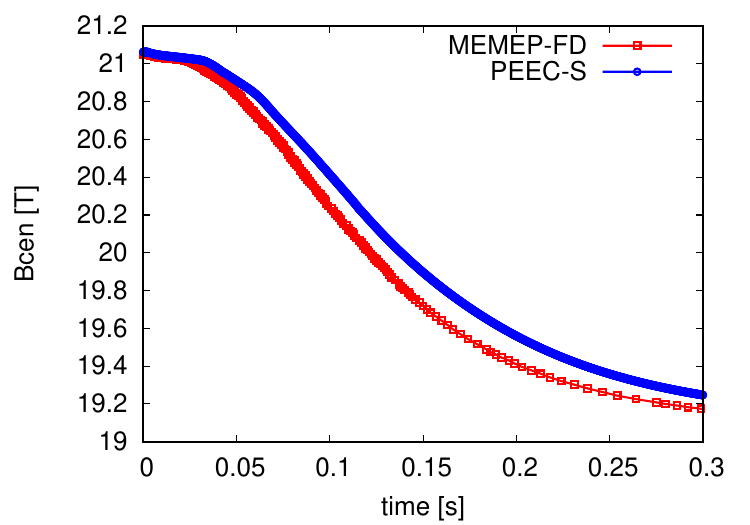}}%
\caption{The modeling results from MEMEP-FD and PEEC-S agree. Computations for a double pancake assuming uniform $J_\varphi$ and $J_r$ at each homogenized turn. Each homogenized turn is made of 5 real turns. \label{f.bench}}%
\end{figure}

We benchmarked the electrothermal model above, MEMEP-FD, with the PEEC-S model. In this comparison, we assumed that the current density is uniform in \R{the cross-section of each homogenized turn}. For MEMEP-FD this assumption is implemented using uniform $J_\varphi$, $J_r$, $\vB$, and $T$ at each homogenized turn, which is made of 5 turns ($n_{\rm nei}$=5). For benchmarking, we used a double pancake model (100 turns per pancake, 25 mm inner radius, Fujikura tape, 10$^{-6}$ \Ohmmm contact resistance, 19 T applied flux density), with a defect introduced at turns 21-25, reducing $J_c$ to 10 \% of its original value exponentially at time constant of 1 ms. \R{Alternatively, we could triger quench by applying heat generated by a local heater, such as when determining the minimum quench energy \cite{haroE2015IES}. However, we prefer reducing $J_c$ in few turns not to} introduce additionnal energy into the system. The current was ramped to 500 A in 500 s, and we impose a 1 V limitation at the terminals of the power source. The models show good agreement for maximum temperature (figure \ref{f.bench}(left)) and magnetic flux density at the bore center \R{(figure \ref{f.bench}(right))} despite small differences due to meshing variations. Homogenized models aligned well with detailed models, significantly reducing computation time \R{(see Appendix \ref{s.A_benchmark} for results of the detailed model)}.

\begin{figure}[tbp]
\includegraphics[trim=0 0 0 0,clip,width=9 cm]{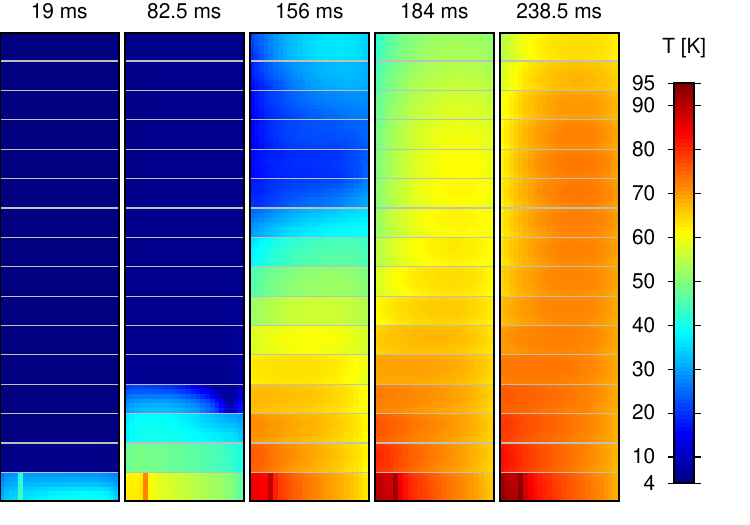}\\%
\includegraphics[trim=0 0 0 0,clip,width=9 cm]{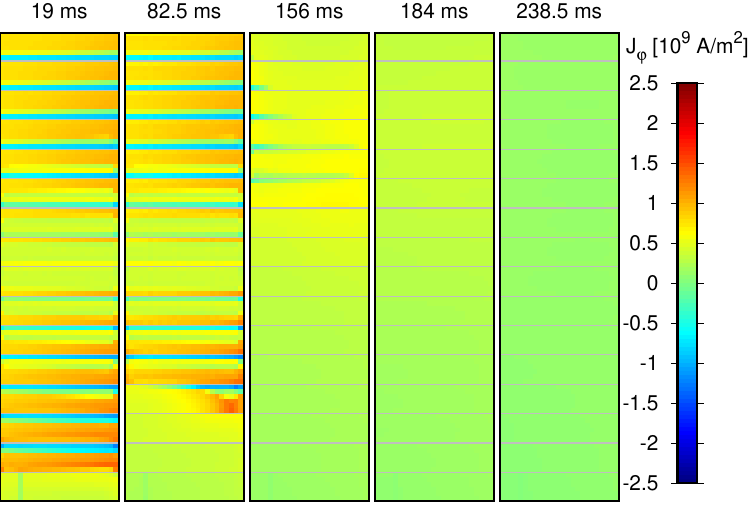}\\%
\includegraphics[trim=0 0 0 0,clip,width=9 cm]{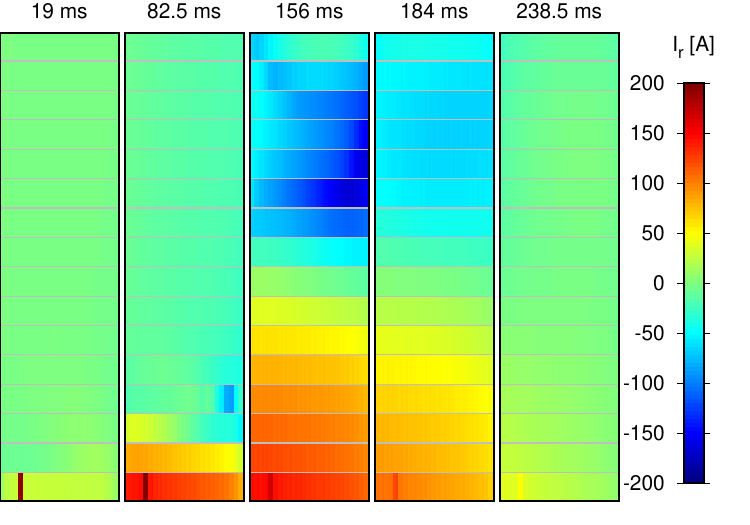}%
\caption{Evolution of the electrothermal quench in the cross-section of the \R{metal-insulated} REBCO insert made of Theva APC tape when there \R{are damaged turns} at the bottom pancake. \R{The damaged turns are 41 to 50 from the inner radius, which consist in one homogenized turn}. Screening currents are taken into account. \R{Quantity $I_r$ is the turn-to-turn radial current.} \label{f.TJ}}%
\end{figure}

\begin{figure}[tbp]
\includegraphics[trim=0 0 0 0,clip,width=9 cm]{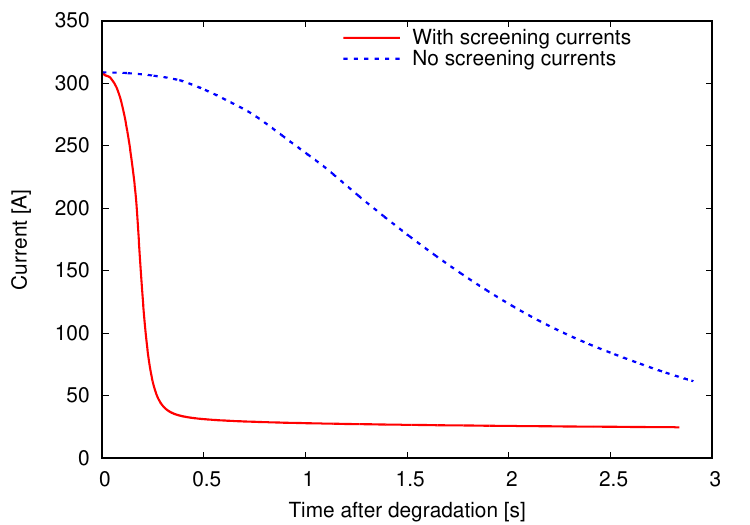}%
\caption{After electrothermal quench, the current decreases due to voltage limiting in the source \R{(1 V)} and decrease of critical current in several turns of the coil \R{below the operation current (see figure \ref{f.TJ})}. \label{f.It}}%
\end{figure}

Next, we apply the MEMEP-FD method to analyze the impact of screening currents on the time evolution of electrothermal quench in the 32 T magnet insert. As quench scenario, we consider that at the middle of the current plateau there appears a damaged superconducting region at turns 41-50 of the bottom pancake, where $J_c(B,\theta,T)$ is reduced to 0.1 times the original. The results are independent on the cause of the damage, which could be due to mechanical stress, \R{among other reasons}. Again, we consider 1 V voltage limitation. \R{After solving $T$, $J_\varphi$ and $I_r$ (figure \ref{f.TJ}),} the \R{net} current decreases with time when electrothermal quench propagates across the magnet (figure \ref{f.It})\R{, due to voltage limitation. Indeed, saturated turns ($J>0$ and $J>J_c$ across the whole turn) present significant voltage. If the accumulated voltage across the whole insert overcomes 1 V, current decreases in order to accommodate the voltage to 1 V. At the beginning, a tiny decrease in $I$ is sufficient to keep the voltage below the imposed limit (see figures \ref{f.Itall} and \ref{f.Vt} of the appendix). This is because the coil large inductance causes that a small current decrease causes inductive voltage that compensates most of the resistive voltage. A substantial amount of the cross-section needs to experience quench in order to achieve a substantial current reduction.}

\begin{figure}[tbp]
\includegraphics[trim=0 0 0 0,clip,width=9 cm]{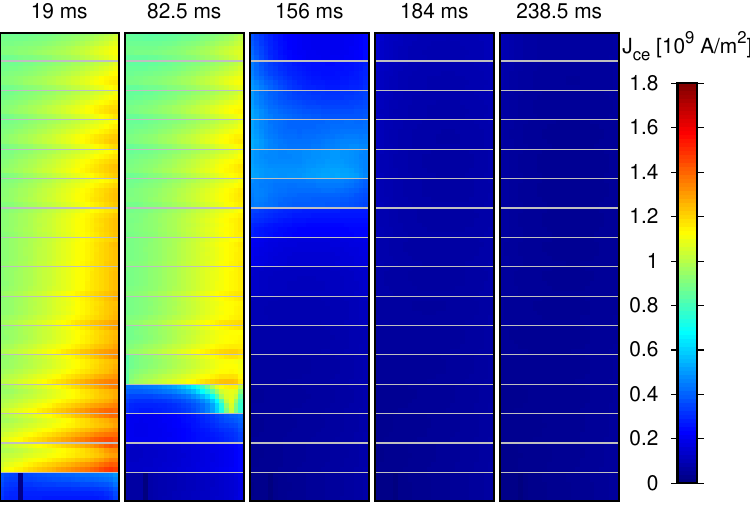}\\%
\includegraphics[trim=0 0 0 0,clip,width=9 cm]{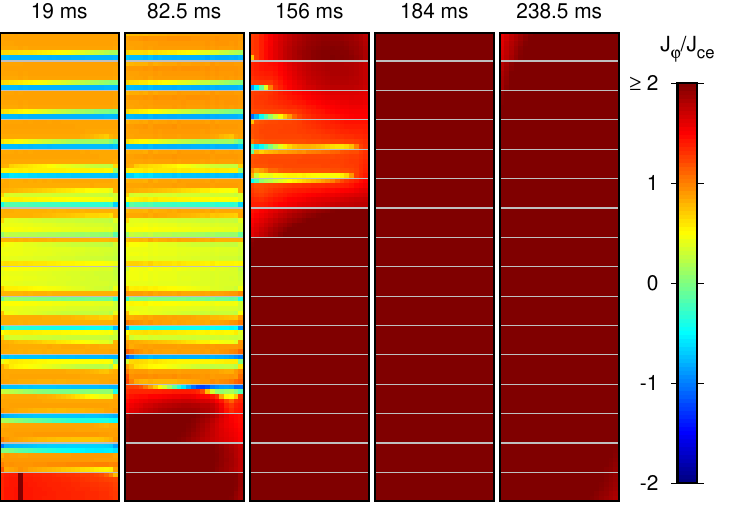}\\%
\includegraphics[trim=0 0 0 0,clip,width=9 cm]{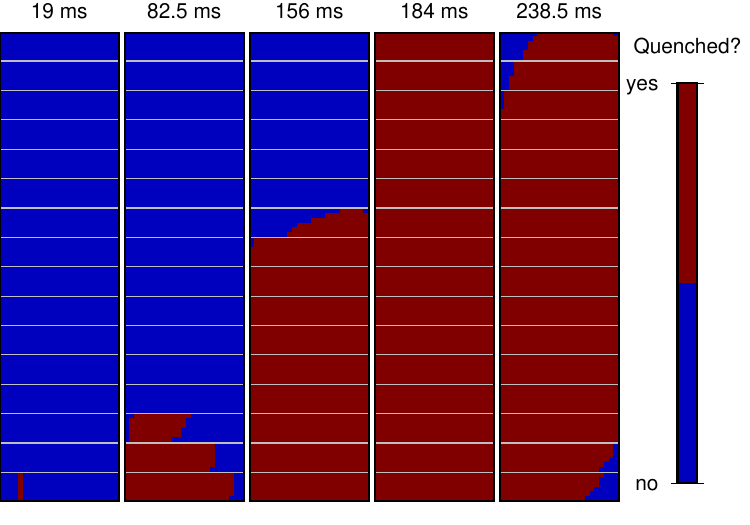}%
\caption{\R{Time evolution of the engineering current density, $J_{ce}$, normalized angular density, $J_\varphi/J_{ce}$, and quenched region. For the quenched region, we use the criterion of $|J_\varphi|/J_{ce}\ge 2$, where the heat generation at the quenched section due to angular current is approximately half of the heat generation in normal state or higher (see text).} \label{f.jJc}}%
\end{figure}

Here, we consider as quenched superconducting regions those where $|J_\varphi|/J_{ce}\ge 2$ as criterion \R{(see section \ref{s.Jquench})}. For REBCO this is a more strict criterion than $T>T_c$ because for high field magnets, there is strong suppression of $J_{ce}$ \R{at temperatures well} below $T_c=92$ K \R{(see figure \ref{f.JcBT}). The quenched zone is shown in figure \ref{f.jJc}.} \R{In that figure we also show the time evolution of $J_{ce}$ and $|J_\varphi|/J_{ce}$ for completeness. The asymmetry of $J_{ce}$ with respect to $z$ at the beginning of the quench is due to the asymmetric $J_c$ of the Theva tape with respect to the magnetic field orientation (see figure \ref{f.JcBT} in the appendix). During quench evolution, $J_{ce}$ decreases due to the local increase of temperature. The modulus of $J_\varphi/J_{ce}$ is around 1 (but slightly below) at the regions where the temperature has not yet increased, and hence the generated power there due to $J_\varphi$ is negligible. Regions with $|J_\varphi|/J_{ce}$ well above 1 result in substantial heat generation.}

\begin{figure}[tbp]
\includegraphics[trim=0 0 0 0,clip,width=9 cm]{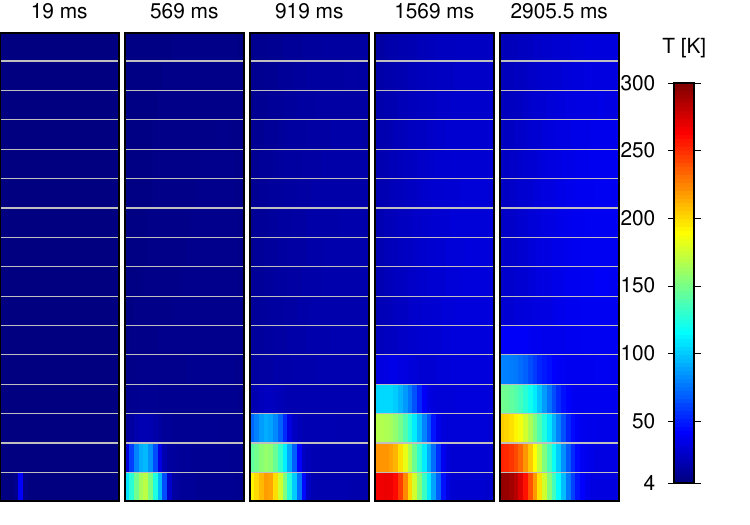}%
\caption{Neglecting screening currents results in artificially slow propagation of electrothermal quench in the REBCO insert of the 32 T magnet (compare with figure \ref{f.TJ}).
\label{f.Tuni}}%
\end{figure}

\begin{figure}[tbp]
{\includegraphics[trim=0 0 0 0,clip,width=10 cm]{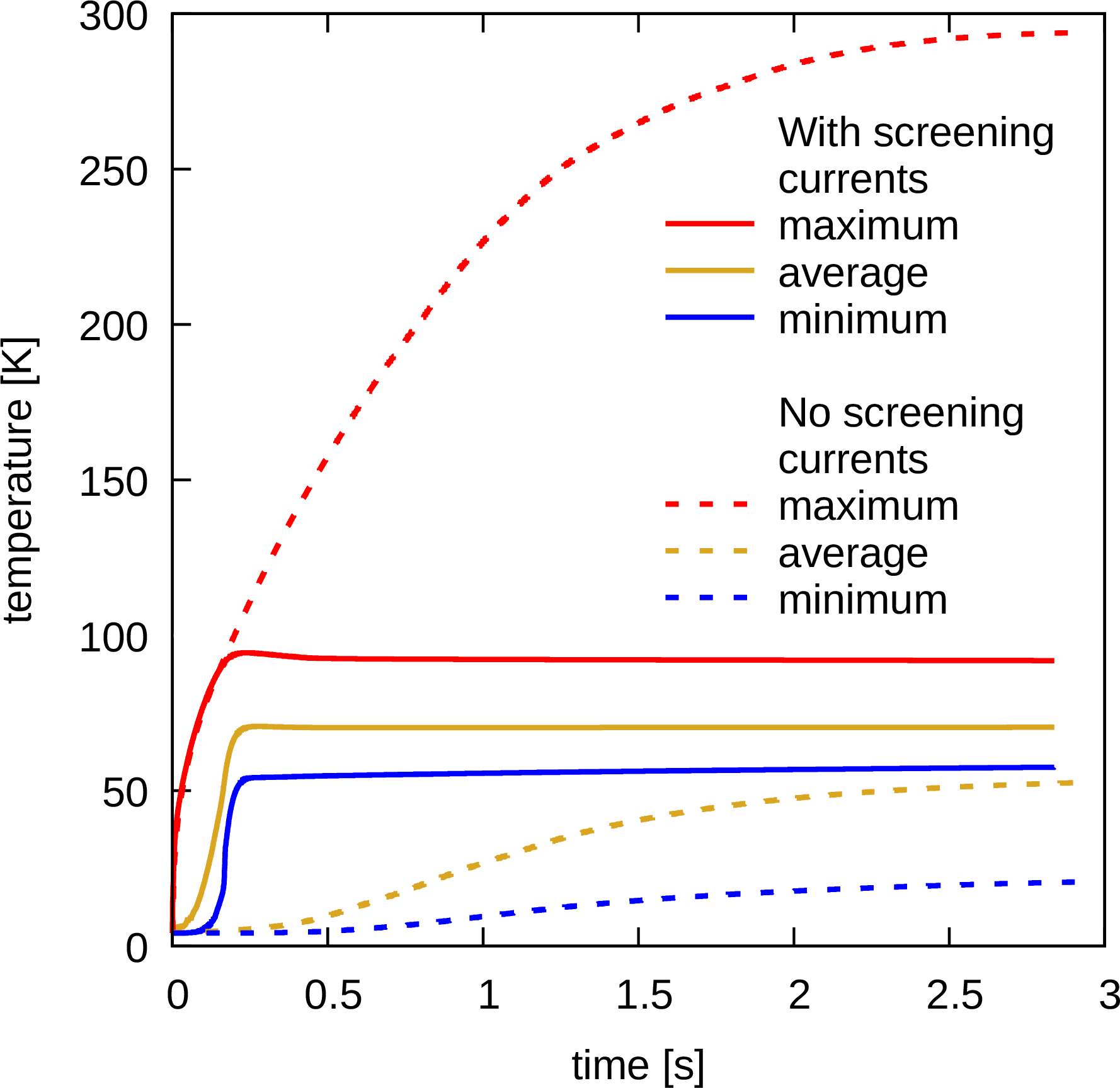}}%
\caption{Minimum, maximum and average temperatures in the REBCO insert with and without screening currents. \label{f.Tav}}%
\end{figure}

When comparing the propagation of the temperature increase with and without screening currents (figures \ref{f.TJ}-\ref{f.Tav}), we see that electrothermal quench propagates much faster with screening currents than without. Indeed, while the whole insert quenches when taking screening currents into account, quench remains localized close to the bottom-left corner when neglecting screening currents (figure \ref{f.Tuni}). The maximum temperature is also much smaller with screening currents (95 K) than witout (around 300 K). Thus, we conclude that screening currents speed-up quench propagation in magnets and they reduce the maximum temperature at quench. \R{Therefore, the assumption of uniform current density produces inaccurate results.}

The causes of fast quench propagation with screening currents are the following. Initially, there is heating on the damaged turns, since there $I$ overcomes $I_c$ (see figure \ref{f.TJ}, 19 ms). Later, the high temperature region propagates (figure \ref{f.TJ}) by three mechanisms. First, thermal conduction transfers heat to neighboring turns and pancakes. Second, a decrease in $J_\varphi$ in one turn causes inductive increase of $|J_r|$ and $J_\varphi$ at other turns. The decrease of $J_\varphi$ is due to either transfer of $J_\varphi$ to $J_r$ at its own turn when $J$ overcomes $J_c$, or by an overall decrease of the net current, $I$. Third, temperature increase by the AC loss caused by a sudden change (usually, suppression) of screening currents. The latter is responsible for fast temperature increase within one pancake, or quench propagation in the radial direction. Indeed, the fast change of screening currents in one turn causes a fast change of the radial magnetic induction, $B_r$, in the neighboring turns, which causes further change in screening currents in these neighboring turns. Quench propagation in the vertical direction is based on the following mechanisms. For a cold unquenched pancake next to a warm quenched one, quench is initiated as follows. First, heat diffuses from the warm pancake and there is also heat generated at the cold pancake by induced radial currents caused by $J_\varphi$ reduction at the warm pancake. This $T$ increase causes a local decrease in $J_c$, which results in an increase of the penetration of screening currents. The latter causes a high local heat generation, which further increases screening current penetration and heating. Then, screening currents also speed-up quench propagation in the axial direction, but at a slower rate than at the radial direction.

\R{Qualitatively, the effect of screening currents is similar to flux jumps \cite{wilson}. A change in magnetic field causes an increase in screening current penetration, which causes AC loss, subsequent temperature increase, and additional screening current penetration and AC loss. This leads to thermal instability, if the magnetic field disturbance is high enough.}

\section{Conclusion}
\label{s.conclusion}

The main conclusion of this work is that screening currents highly increase the quench propagation speed in ultra-high-field magnets. Then, realistic electrothermal modelling should take screening currents into account. Assuming uniform angular current density, $J_\varphi$ at the superconductor will predict substantially slower quench propagation and higher maximum temperature. On one hand, designs based on uniform $J_\varphi$ models are more conservative regarding maximum temperature. On the other, there might be issues in predicting quench in the LTS due to quench in the insert, since the voltage in the LTS due to the decrease in flux of the insert will be under-estimated. In addition, enhanced quench propagation speed due to screening currents could be exploited for future quench-resilient magnet designs.

Our findings will have an impact in the design of ultra-high-field magnets for NMR or user facilities, and possibly for other kinds of magnets like those for fusion energy.

\section*{Acknowlegements}

We acknowledge Oxford Instruments, and in particular A Varney, S Boll and A Davydov for details on the LTS magnet geometry and discussions. This project has received funding from the European Union's Horizon 2020 research and innovation programme under grant agreement No 951714 (superEMFL), and the Slovak Republic from projects APVV-24-0654 and VEGA 2/0098/24. Also funded by the EU NextGenerationEU through the Recovery and Resilience Plan for Slovakia under the project No. 09I04-03-V02-00039. Any dissemination of results reflects only the authors' view and the European Commission is not responsible for any use that may be made of the information it contains.

\section*{Declaration}

No artificial intelligence has been used in writing this article or preparing any graph.

\newpage

\newpage

\appendix

\section{Material properties}

The superconducting tapes consist on several parallel layers: REBCO superconductor, hastelloy, copper, and silver, as well as Durnomag as metal-insulation between turns of each pancake. The total thicknesses of each material is in table \ref{t.insert}. Here, we assumed that the material properties of silver are the same as copper, for simplicity. The reason is that the total copper thickness (18 $\mu$m) is much higher than silver (2 $\mu$m) and their electric properties are within the same order of magnitude. In this article, we also assume that the properties of the Hastelloy and Durnomag metals is the same as stainless steel. The reasons are that, firt, their electric and thermal properties are similar and, sencond, the electric, thermal and mechanical properties of stainless steel are well documented at a wide range of temperatures \cite{NISTmaterials}, while that is not the case for Hastelloy and Durnomag. In addition, we assume that the electric and thermal properties of REBCO are the same as stainless steel. The reason is that the thickness of REBCO is much smaller than the Hastelloy and Durnomag combined, and that they are of the same order of magnitude as REBCO. As in \cite{dadhichA2024SSTa}, we consider that the pancakes are separated by a G10 insulating layer of 500 $\mu$m thickness.

In this work, we consider homogenized properties of the tape from the properties of each layer, as detailed in section \ref{s.method} and in \cite{pardoE2023IES}.

\subsection{Electric properties}

The assumed relation between the electic field, $\vE$, and current density, $\vJ$, are given by equations (\ref{JE}) and (\ref{JrEr}). The key parameters are the resistivities of the metals ($\rho_{{\rm n}m}$), the engineering critical current density of the superconductor ($J_{ce}$), and its power-law exponent ($n$). For the metals, we take the dependence of the resistivity on temperature and magnetic field from \cite{NISTmaterials}. For copper, we consider RRR=100. For the superconductor, we assume a power-law exponent of $n$=30.

\begin{figure}[tbp]
{\includegraphics[trim=50 0 55 9,clip,height=8.5 cm]{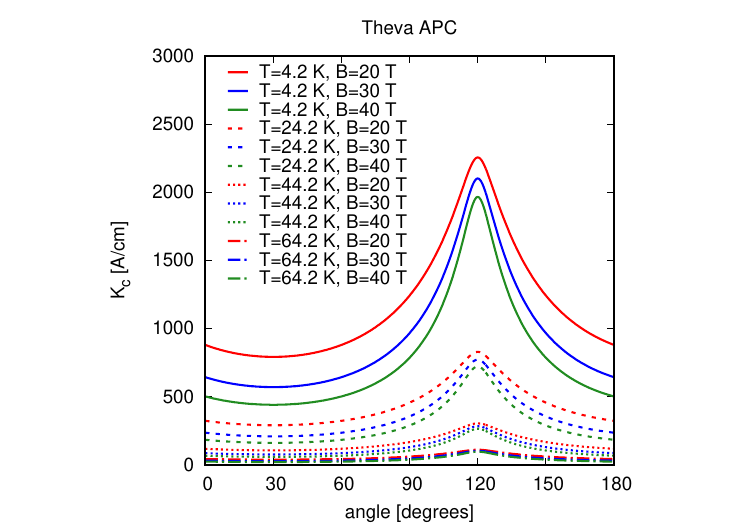}}%
{\includegraphics[trim=95 0 55 9,clip,height=8.5 cm]{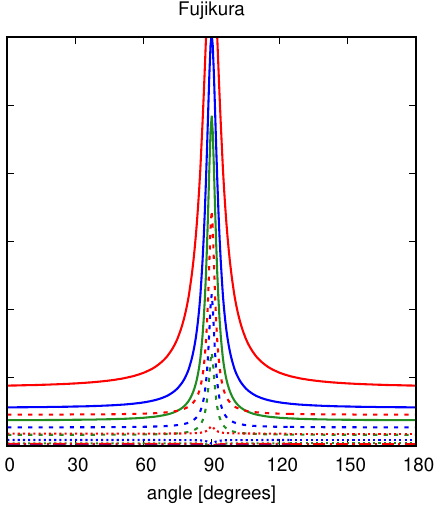}}%
\caption{The superconducting Theva APC and Fujikura tapes present a different angular dependence of the tape critical current per unit tape width, $K_c$. The shown dependences on angle, temperature and magnetic field are from the fits in (\ref{KcTheva}) and (\ref{KcFujikura}), which are based on experimental data. 
\label{f.JcBT}}%
\end{figure}

Next, we present the inputs of critical current per unit tape width, $K_c$, instead of critical current density, $J_c$, since $K_c$ is independent on the assumed thickness of the superconducting layer, $d$. Both quantities are related to each other as $J_c=K_c/d$. If we take $d$ as the total tape thickness, $J_c$ becomes the engineering current density, $J_{ce}$. 

The critical current per unit tape width of the Theva APC tape is taken from the following fit of the experimental data in \cite{senatoreC2024SST}. The fit applies for $T<T_c=92$ K and $B\ge 10$ T,
\begin{equation} \label{KcTheva}
K_c=\frac{K_{c0}e^{-(T-T_l)/T_{\rm ch}}}{\left ( 1+\frac{\sqrt{(kB\sin(\theta-\theta_0))^2+(B\cos(\theta-\theta_0))^2}}{B_{cc}} \right )^{b}},
\end{equation}
where the constants above are $K_{c0}=2.638\cdot 10^{5}$ A/m = 2638 A/cm, $T_l=4.2$ K, $T_{
\rm ch}=20$ K, $k=0.08$, $b=1.188$, $\theta_0=\pi/6$ rad = 30$^{\rm o}$, $B_{cc}=11.44$ T. Above, the angle $\theta$ is defined by $\theta={\rm atan2}(B_z,B_r)$, which is 0 when $\vB$ is perpendicular to the tape surface and $\pi/2$ when $\vB$ is parallel. For $T\ge T_c$ the superconductor becomes normal, and hence $K_c=0$. 

For the Fujikura tape we use the fit based on experimental data from \cite{fleiterJ2014CERN}, which we write in the compact form
\begin{equation} \label{KcFujikura}
K_c(B,\theta,T)=K_{cc}(B,T)+\frac{K_{cab}(B,T)-K_{cc}(B,T)}{1+\left ( \frac{\theta'}{g(T,B)} \right )^\nu}
\end{equation}
with
\begin{eqnarray}
g(T,B) & = & g_0+g_1\exp \left \{ -g_2  \exp(g_3T) B' \right \} \nonumber \\
B' & = &	\begin{cases} 
					B & \text{if $B\ge B_{\rm min}$ }\\
					B_{\rm min} & \text{if $B<B_{\rm min}$}
					\end{cases} \nonumber \\
\theta' & = & \left | {\rm atan}\frac{B_r}{B_z} \right | = \frac{\pi}{2}-\theta \label{thp}
\end{eqnarray}
and
\begin{eqnarray}
K_{cc}(B,T) & = &	
\begin{cases}									
\frac{\alpha_c}{B'}\left ( \frac{B'}{B_{ic}} \right )^{p_c} 
\left [ 1- \left ( \frac{B'}{B_{ic}} \right ) \right ]^{q_c}f(T)^{\gamma_c} &
\text{if $B'<B_{ic}$ } \\
0 & \text{if $B'\ge B_{ic}$ } 
\end{cases} \nonumber \\
f(T) & = & 1-\left ( \frac{T}{T_{c0}} \right )^n \nonumber \\
B_{ic} & = & B_{i0c}f(T)
\end{eqnarray}
and as well
\begin{eqnarray}
K_{cab}(B,T) & = & 
\begin{cases}
\frac{\alpha_{ab}}{B'} \left ( \frac{B'}{B_{iab}} \right )^{p_{ab}} 
\left [ 1- \left ( \frac{B'}{B_{iab}} \right ) \right ]^{q_{ab}}h(T)^{\gamma_{ab}}
& \text{if $B'<B_{iab}$} \\
0 & \text{if $B'\ge B_{iab}$}
\end{cases} \nonumber \\
h(T) & = & \left [ 1-\left ( \frac{T}{T_{c0}} \right )^{n_1} \right ]^{n_2} + a f(T) \nonumber \\
B_{iab} & = & B_{i0ab}h(T).
\end{eqnarray}
The constants above are $\nu=1.85$, $g_0=0.03$, $g_1=0.25$, $g_2=0.06$ T$^{-1}$, $g_3=0.058$ K$^{-1}$, $B_{\rm min}=0.01$ T, $\alpha_c=3.72\cdot 10^{6}$ AT/m, $p_c=0.5$, $q_c=2.5$, $\gamma_c=2.44$, $T_{c0}=93$ K, $n=1$, $B_{i0c}=140$ T, $\alpha_{ab}=1.366\cdot 10^{8}$ AT/m, $p_{ab}=1$, $q_{ab}=5$, $\gamma_{ab}=1.63$, $n_1=1.4$, $n_2=4.45$, $a=0.1$, $B_{i0ab}=250$ T. Equations (\ref{KcFujikura}) and (\ref{thp}) are valid for $\theta \in [0,\pi/2]$ and are assumed symmetric \R{dependence} beyond this angle range.

\subsection{Thermal properties}

Here, we consider the heat capacity per unit volume, $c_p$, and heat conductivity in the $r$ and $z$ direction, $k_r$ and $k_z$, as the homogenized equivalents, as detailed in section \ref{s.method}. However, the model takes a different thermal properties for the spacer between pancakes, which is made of G10 epoxy-fiberglass. All material properties are the same as in \cite{dadhichA2024SSTa}, which are taken from \cite{NISTmaterials}. However, we consider that the radial thermal resistivity is divided by a certain factor in order to take non-ideal interface thermal conductivity into account. We took a factor 5, in correspondence with thermal conductivity measurements \cite{durochatM2024IES}.

\section{The MEMEP-FD electrothermal model}

\subsection{Minimum Electromagnetic Entropy Production (MEMEP)}

We solve the electromagnetic properties by the Minimum Electro Magnetic Entropy Production (MEMEP) method \cite{pardoE2017JCP}. This method can also take non-insulated and metal-insulated coils into account \cite{pardoE2024SSTa}. As done in \cite{pardoE2024SSTa}, we can take radial currents between turns into account while maintaining axial symmetry. \R{MEMEP is a variational method that, as other variational methods, can efficiently solve highly non-linear $\vE-\vJ$ relations \cite{pardoE2017JCP}. Indeed, we have been able to model superconductors with a power law exponent as high as 1000 \cite{pardoE2017JCP}. Other variational methods have modeled superconductors with the sharp multi-valued $\vE-\vJ$ relation of the Critical State Model, and hence zero resistivity for $|\vJ|<J_c$ and infinite slope of the $\vE-\vJ$ relation at $|\vJ|=J_c$ \cite{bossavitA1994IEM, prigozhinL1997IES, pardoE2012SSTb}.}

In general, MEMEP solves the following equation for the current density as a function of position, $\vr$, and time, $t$, $\vJ(\vr,t)$:
\begin{equation} \label{EJAphi}
\vE(\vJ)=-\frac{\partial\vA[\vJ]}{\partial t}-\frac{\partial\vA_a}{\partial t}-\nabla\phi,
\end{equation}
where $\vE(\vJ)$ is the (non-linear) constitutive relation between the electric field and current density from the material, $\vA[\vJ]$ is the vector potential in Coulomb's gauge ($\nabla\cdot\vA=0$ and $|\vA|\to 0$ at $|\vr|\to \infty$) generated by $\vJ$, which is
\begin{equation} \label{AJ}
\vA[\vJ](\vr,t)=\frac{\mu_0}{4\pi}\int_\Omega \dvol'\frac{\vJ(\vr',t)}{|\vr-\vr'|},
\end{equation}
where $\Omega$ is the region where $\vJ$ is non-zero. In (\ref{EJAphi}), $\vA_a$ is the vector potential in Coulomb's gauge generated by known external sources (or applied vector potential) and $\phi$ is the electrostatic potential. We discretize equation (\ref{EJAphi}) in time as
\begin{equation}
\vE(\vJ)=-\frac{\vA[\Delta\vJ]}{\Delta t}-\frac{\Delta \vA_a}{\Delta t}-\nabla\phi,
\end{equation}
where $\Delta t$ is a certain finite time step, $\Delta\vJ(t)\equiv \vJ(t)-\vJ(t-\Delta t)$ and $\Delta \vA_a\equiv \vA_a(t)-\vA_a(t-\Delta t)$. For a known $\vJ(t-\Delta t)$, solving the above equation is the same as minimizing the following functional with respect to $\Delta \vJ$
\begin{eqnarray} \label{F3D}
F[\Delta\vJ] & = & \int_\Omega\dvol\left \{ \frac{1}{2}\Delta\vJ \cdot \frac{\vA[\Delta\vJ]}{\Delta t}  + \Delta\vJ \cdot \frac{\Delta\vA_a}{\Delta t} + U(\Delta\vJ+\vJ_p) \right \} \nonumber \\
& & + \int_\Omega\dvol	 (\vJ_p+\Delta\vJ)\cdot\nabla\phi, 
\end{eqnarray}
where $\vJ_p\equiv \vJ(t-\Delta t)$ and
\begin{equation}
U(\vJ)\equiv \int_0^\vJ\dif\vJ'\cdot\vE(\vJ').
\end{equation}
We name $U(\vJ)$ above as dissipation factor. The minimum of this functional is unique for each time step and the Euler equation of this functional is (\ref{F3D}) \cite{pardoE2017JCP}. The time evolution for any desired sequence of times can be obtained by assuming the initial condition of field cool; which is $\vJ(t=0)=0$ and $\vA_a(t=0)=0$. 

If the current leads extend far away from the coil magnetic field and they are straigth close to their ends, the last term in (\ref{F3D}) becomes $-VI$, where $V$ is the voltage and $I$ is the input currrent. The reasons are the following. From vector calculus, $\nabla\phi \cdot \vJ =\nabla\cdot(\phi\vJ)-\phi\nabla\cdot\vJ$. Since $\nabla\cdot\vJ=0$, the last term in (\ref{F3D}) becomes $\int_{\partial\Omega}\dif s \phi{\bf n}\cdot\vJ$, where ${\bf n}$ is the unit vector normal to the cross-section. If there is no component of $\vB$ along the lead transport direction, $\vJ$ follows the transport direction, and hence it is parallel to ${\bf n}$. In that case, $\vE$ is also parallel to $\vn$ because all materials in the tape are isotropic. Since the leads are not under the influence of the coil magnetic field, in Coulomb's gauge $\vA=\vA[\vJ]$ and hence $\vA\|\vJ$. Then, $\vA\|\vn$. From $\vE=-\partial\vA/\partial t - \nabla\phi$, we obtain that $\nabla\phi\|\vn$. Therefore, $\nabla\phi=\partial_{\vn}\phi\vn$, where $\partial_{\vn}\phi$ is the spacial derivative along the normal of the tape cross-section. Then, $\phi$ does not depend on the other two coordinates that are within the tape section. Therefore, $\phi$ is uniform at the tape section on the current leads. Then the last term in (\ref{F3D}) becomes
\begin{eqnarray}
\int_\Omega\dvol \nabla\phi\cdot\vJ & = & \int_{\partial\Omega}\dif s\vn\cdot(\phi\vJ) = 
\int_{S_{\rm out}}\dif s\vn\cdot\vJ\phi-\int_{S_{\rm in}}\dif s\vn\cdot\vJ\phi \nonumber \\
& = & (\phi_{\rm out}-\phi_{\rm in})I=-VI,
\end{eqnarray}
where $S_{\rm out}$ and $S_{\rm in}$ are the cross-sections of the output and input terminals, respectively, and $\phi_{\rm out}$ and $\phi_{\rm in}$ are their output and input electrostatic potentials.	When using current constraints, $I$ is fixed, and hence the functional is independent on the specific current distribution. In that case, the therm $-VI$ could be dropped from the functional. However, when using voltage constraints, the current $I$ can vary, and hence the term $-VI$ should remain.

Next, we assume axial symmetry. As detailed in \cite{pardoE2024SSTa}, we can use axial symmetry to model metal-insulated coils by imposing the following current conservation condition in each turn $i$
\begin{equation}
I_{\varphi i}+I_{r i}=I,
\end{equation}
where $I_{\varphi i}$ and $I_{ri}$ are the total current at turn $i$ in the angular and radial directions, respectively. When considering angular symmetry and neglecting the vector potential from the radial currents, the functional in (\ref{F3D}) becomes
\begin{equation} \label{FJcyl}
F[\vJ]=2\pi\int_{\Omega_s}\dif r\dif z\ r \left \{ \half \Delta J_\varphi \frac{A_\varphi [\Delta J_\varphi]}{\Delta t} + \Delta J_\varphi \frac{\Delta A_{a\varphi}}{\Delta t} + U(\vJ)
\right \} -VI, 
\end{equation}
where $\Omega_s$ is the cross-sectional region of the conducting or superconducting material. Since in our case $E_\varphi$ only depends on $J_\varphi$ [see equation (\ref{JE})] and $E_r$ depends only on $J_r$ [see equation (\ref{JrEr})], the dissipation factor $U(\vJ)$ becomes
\begin{eqnarray}
&& U(\vJ)=U_\varphi(J_\varphi)+U_r(J_r) \\
&& U(J_\varphi)=\int_0^{J_\varphi}\dif J_\varphi E_\varphi(J_\varphi) = E_\varphi(J_\varphi)J_\varphi-\int_0^{E_\varphi(J_\varphi)}\dif E_\varphi' J_\varphi(E_\varphi')
\label{Uphi} \\
&& U(J_r)=\int_0^{J_r}\dif J_r' E_r(J_r') \label{Ur}.
\end{eqnarray}
The second part of (\ref{Uphi}) is useful when the constitutive relation is in form of $J_\varphi(E_\varphi)$, as in (\ref{JE}). Integrating (\ref{Uphi}) with (\ref{JE}) and (\ref{Ur}) with (\ref{JrEr}), we find
\begin{eqnarray}
&& U_\varphi(J_\varphi)=E_\varphi(J_\varphi)J_\varphi - \frac{1}{2\rho_n}E_\varphi(J_\varphi)^2 - \frac{J_{ce}}{1/n+1} \left | \frac{E_\varphi(J_\varphi)}{E_c} \right |^{\frac{1}{n}} |E_\varphi(J_\varphi)| \\
&& U_r(J_r)=\frac{1}{2d}J_r^2R_s,
\end{eqnarray}
where $E_\varphi(J_\varphi)$ is obtained from a given $J_\varphi$ by numerically inverting (\ref{JE}).

For a given input current $I$, we solve $J_r$ and $J_\varphi$ by means of the algorithm detailed in \cite{pardoE2024SSTa}. In this work, we consider voltage limitation as follows. First, we solve $J_\varphi$ and $J_r$ according to the desired $I$ of the magnet. Then, we calculate $V$ from the radial current as
\begin{equation}
V=\sum_{i=1}^N V_i=\sum_{i=1}^N \frac{I_{ri}S_{ri}}{R_{si}},
\end{equation}
where $V_i$ is the voltage drop in turn $i$, $N$ is the total number of turns in the insert, $S_{ri}$ is the turn surface perpendicular to the $r$ direction and $R_{si}$ is the contact resistance at turn $i$ ($R_{si}=R_s$ in our case). If $V$ is below and equal to the limiting voltage, $V_{\rm lim}$, the computation ends. Otherwise, we modify $I$ (usually, we reduce $I$) until the new voltage becomes the input voltage within a certain tolerance ($\pm 1$ \% in our case).

\subsection{Thermal modeling by finite differences}
\label{s.FD}

For thermal modelling, we use a method based on finite differences (FD), as detailed in \cite{dadhichA2024SSTa}. This method solves the thermal diffusion equation
\begin{equation} \label{Tdiff}
 { c_p(T) \frac{\partial T}{\partial t} = \nabla \cdot \left ( \ten{k}(T) \nabla T \right ) + p   },
\end{equation}
where $c_p$ is the heat capacity per unit volume at constant pressure, $c_p=\rho_mC_p$ with $C_p$ being the specific heat capacity and $\rho_m$ the mass density; $\ten{k}$ is the thermal conductivity tensor; and $p$ is the heat power density, which is $p=\vJ\cdot\vE$. In cylindrical coordinates, the thermal diffusion equation becomes
\begin{equation} \label{Tdiffcyl}
c_p(T) \frac{\partial T}{\partial t} =   \frac{1}{r} \frac{\partial }{\partial r} \left (r \cdot k_r(T) \frac{\partial T}{\partial r} \right ) + \frac{\partial }{\partial z} \left ( k_z(T) \frac{\partial T}{\partial z} \right ) + p.
\end{equation}
The numerical method takes the dependence of $c_p$, $k_r$, and $k_z$ into account, as well as any possible property non-homogeneities or discontinuities. We solve the partial differential equation of (\ref{Tdiffcyl}) by discretizing the problem in space and time, as detailed in \cite{dadhichA2024SSTa}.

The electromagnetic, MEMEP, and thermal, FD, methods are coupled iteratively \cite{dadhichA2024SSTa}, as follows. MEMEP finds $J_\varphi$ and $J_r$. From these, we calculate $p=\vJ\cdot\vE$. Then, we solve $T$ by FD. Afterwards, we update $\rho_{\rm n}$ and $J_{ce}$ according to the new local temperature. Finally, we iterate until the maximum temperature difference between two iterations is below a certain tolerance (100 mK in this work). Since accurate FD computations require small time steps, we enable to use smaller time steps for FD than MEMEP. For this, we assume that $p=\vJ\cdot\vE$ is constant in the MEMEP time step. The time step for FD is determined by the stability condition \cite{dadhichA2024SSTa} 
\begin{equation} \label{stabilityFD}
\Delta t \le \frac{1}{2} \frac{\Delta r^2 \Delta z^2}{\Delta r^2 + \Delta z^2} \cdot \frac{{\rm min} (c_p)}{{\rm max} (k)},  
\end{equation}
wgere ${\rm min}(c_p)$ is the \R{minimum} value across the insert and ${\rm max}(k)$ is the maximum $k_\varphi$ or $k_r$. If (\ref{stabilityFD}) provides a value larger than $\Delta t$ of MEMEP, $\Delta t$ is the same as $\Delta t$ for MEMEP.

\section{Benchmark between MEMEP-FD and PEEC-S}
\label{s.A_benchmark}

\begin{figure}[tbp]
{\includegraphics[trim=0 0 0 0,clip,width=8.2 cm]{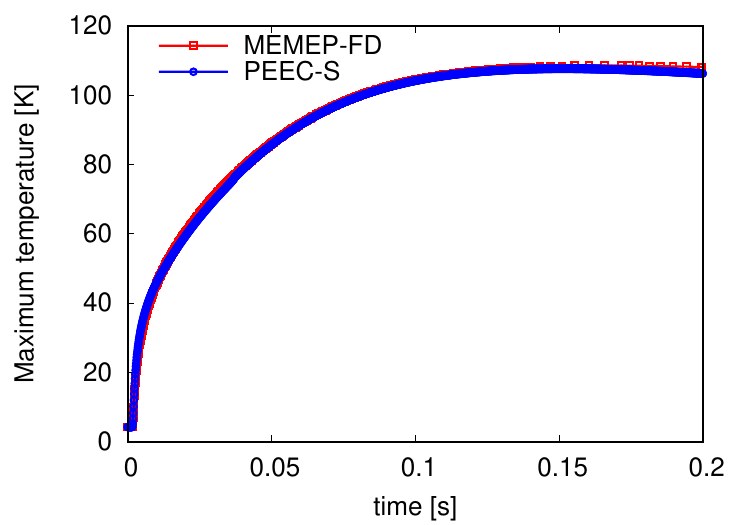}}%
{\includegraphics[trim=0 0 0 0,clip,width=8.2 cm]{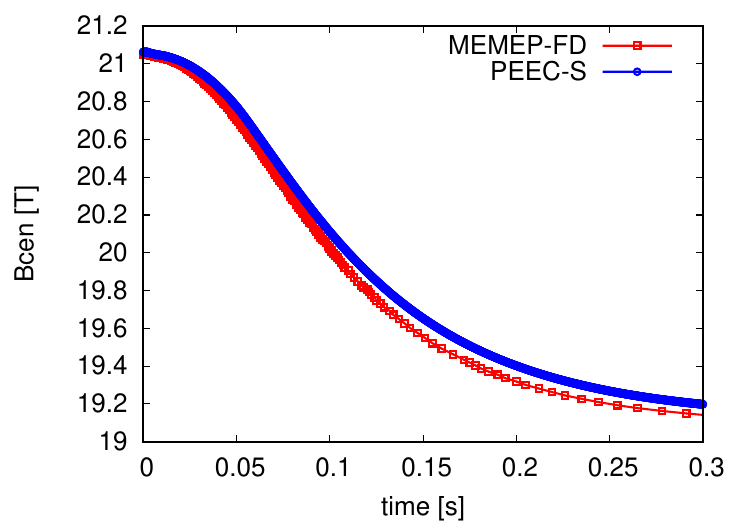}}\\%
{\includegraphics[trim=0 0 0 0,clip,width=8.2 cm]{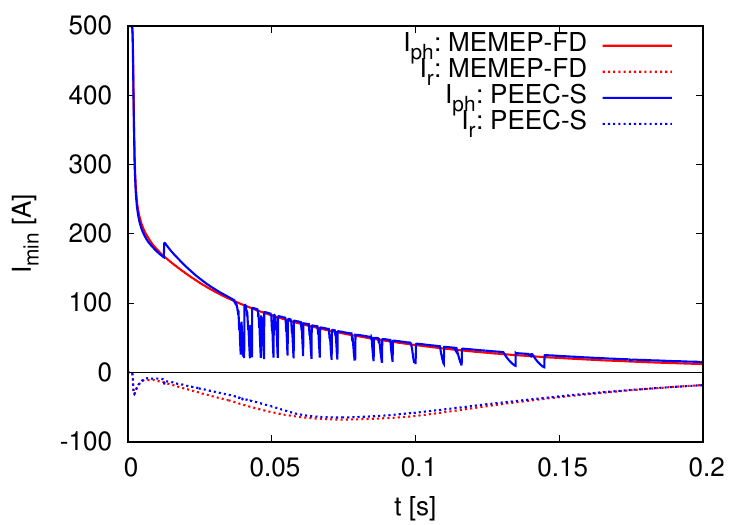}}%
\caption{The modelling results from MEMEP-FD and PEEC-S for a double pancake agree. Computations consider each turn independently, but with homogenized material properties. Screening currents are neglected. \label{f.benchDet}}%
\end{figure}

As described in section \ref{s.results}, we benchmark the computations from MEMEP-FD with PEEC-S for a double pancake coil with a defect in 5 turns. In the benchmark, we neglected screening currents because in the PEEC-S implementation in this work, it did not take screening currents into account. Then, we assume that both angular and radial current densities, $J_\varphi$ and $J_r$, are uniform \R{at each homogenized turn}. Thus, we can define the angular and radial currents as $I_\varphi=J_\varphi S$ and $I_r=J_rS$, where $S$ is the cross-section of the double pancake. We performed benchmarking for both the detailed model, where we consider each turn separately, and the homogenized approach, where we group neighboring turns in groups of 5 and we assume that $J_\varphi$ and $J_r$ are the same in all 5 turns for a given axial coordinate $z$.

For the detailed model, the maximum temperature and magnetic field at the bore center, $B_{\rm cen}$, for MEMEP-FD and PEEC-S agree with each other (figure \ref{f.benchDet}). The main behavior of the minimum current current across the coil section in the angular direction, $I_{\varphi,{\rm min}}$, and in the radial direction, $I_{r,{\rm min}}$, agree. The sharp local minima that appear for PEEC-S are due to numerical error. The value of $I_{\varphi,{\rm min}}$ is sensitive to numerical error because $I_{\varphi,{\rm min}}$ occurs in the few damaged turns, which are highly localized. However, the results of most practical importance (maximum temperature and $B_{\rm cen}$) are free from these numerical errors. The remaining discrepancy is due to different meshing in the thermal models: the PEEC-S model assumes a single element per turn, while the MEMEP-FD model assumes 6 elements (in the $z$ direction). In addition, MEMEP-FD considers 2 elements in the spacer between pancakes (made of G10), while PEEC-S assumes only 1 element. After solving the temperature, the MEMEP-FD model averages this quantity over each turn cross-section in order to obtain uniform $J_c$.

For the homogenized approach, the results from MEMEP-FD and PEEC-S agree (figure \ref{f.bench}). In addition, there is also agreement between the detailed and homogenized models. Therefore, we can use the homogenized approach in order to obtain useful results but by using lower computing time.

Benchmarking enabled improvements in MEMEP-FD, such as the importance of considering the homogenized resistivity in the angular direction accurately (see section \ref{s.method} and \cite{pardoE2023IES}), rather than the approximation of \cite{pardoE2024SSTa} and \cite{dadhichA2024SSTb}. This improvement reduces the heat generation by at least 20 \%.

\section{Time evolution during quench for Theva and Fujikura tapes}

\begin{figure}[tbp]
{\includegraphics[trim=0 0 0 0,clip,height=7.5 cm]{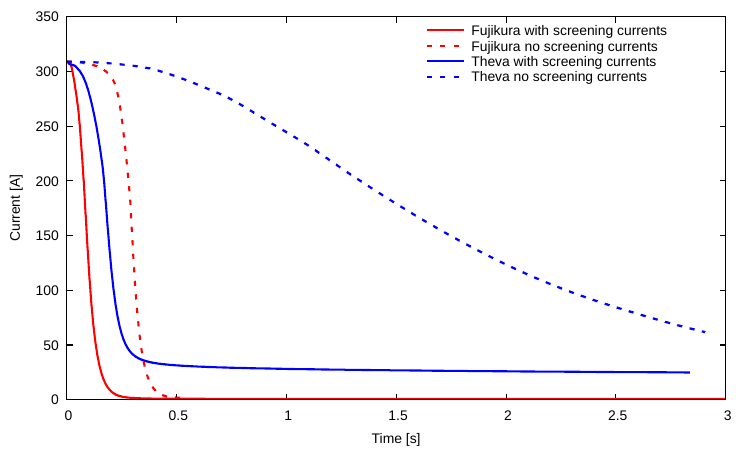}}%
\caption{The decay of current due to quench for 1 V voltage limitation is faster for the insert with Fujikura tape than for the Theva one. Screening currents speed up the decay in both cases. \label{f.Itall}}%
\end{figure}

\begin{figure}[tbp]
{\includegraphics[trim=0 0 0 0,clip,width=16.4 cm]{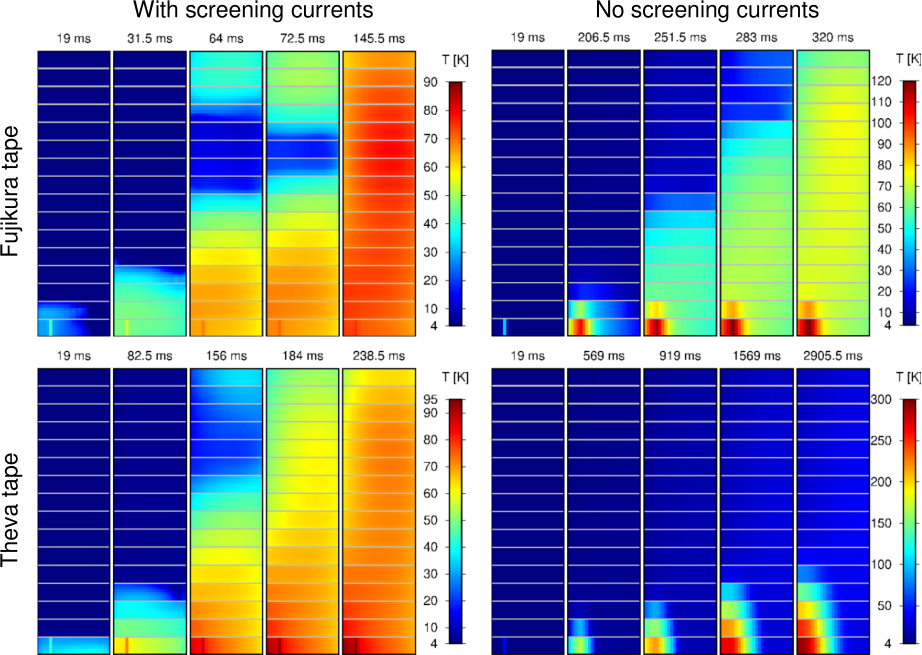}}%
\caption{Time evolution of the temperature in the cross-section of the REBCO insert (time starting from the appearance of the defect close to the bottom-left corner). Screening currents highly increase the propagation of the temperature in the insert for both Fujikura and Theva tapes. \label{f.Tall}}%
\end{figure}

\begin{figure}[tbp]
{\includegraphics[trim=0 0 0 0,clip,width=16.4 cm]{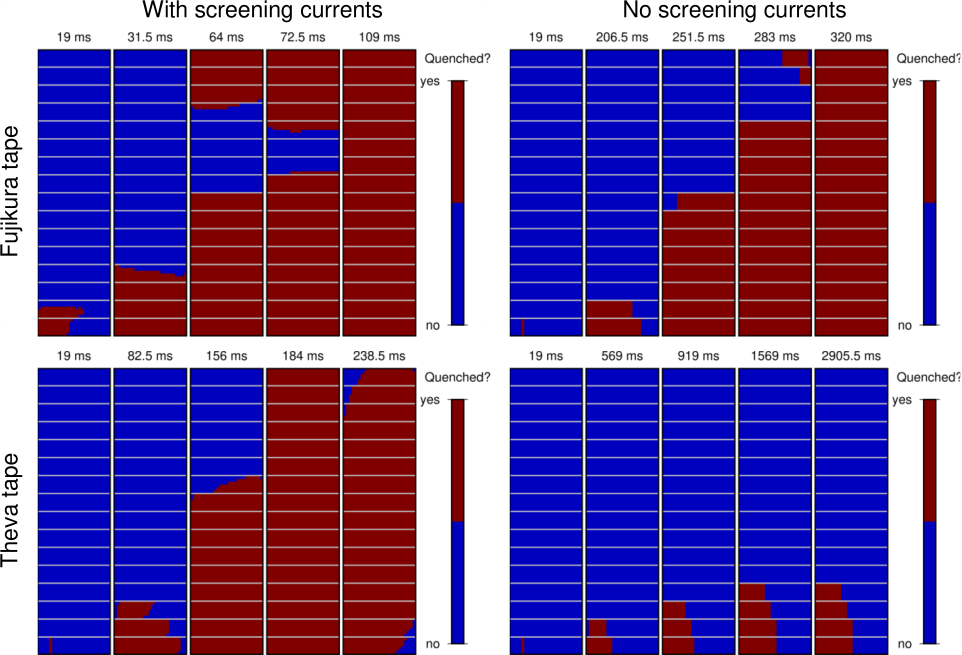}}%
\caption{Time evolution of quench propagation in the REBCO insert taking as quench criterion of \R{$|J_\varphi|/J_c \ge 2$}. Screening currents dramatically highly speed-up quench propagation, especially for the Theva tape. \label{f.quenchj}}%
\end{figure}

\begin{figure}[tbp]
{\includegraphics[trim=0 0 0 0,clip,width=16.4 cm]{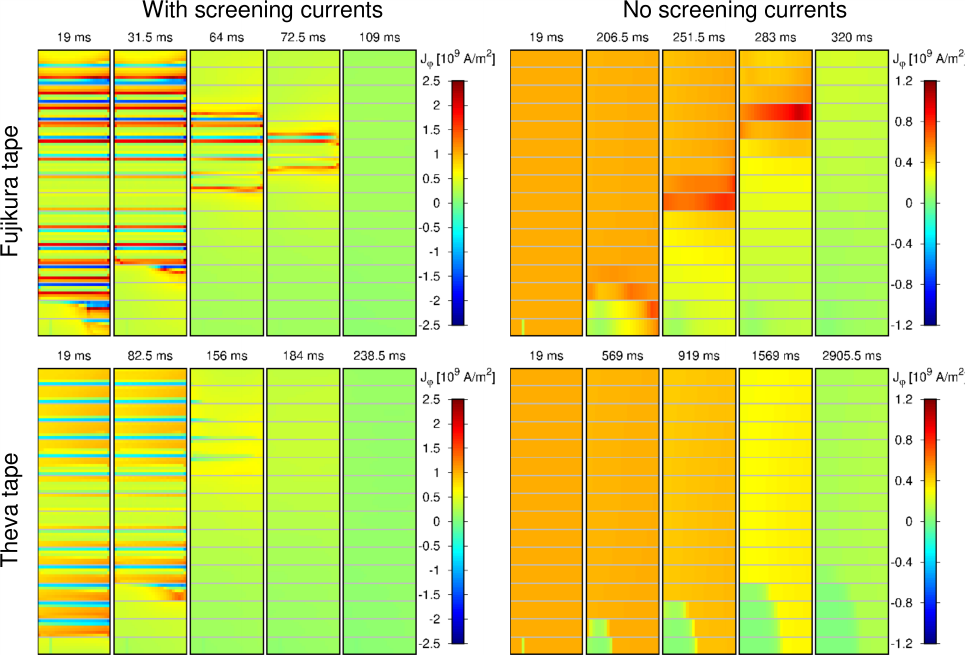}}%
\caption{Screening currents appear in the time evolution of the angular current, $J_\varphi$, in the regions that did not yet experience electrothermal quench (left maps). The right maps assume no screening currents, and hence uniform $J_\varphi$ in each turn. 
\label{f.Jphall}}%
\end{figure}

\begin{figure}[tbp]
{\includegraphics[trim=0 0 0 0,clip,width=16.4 cm]{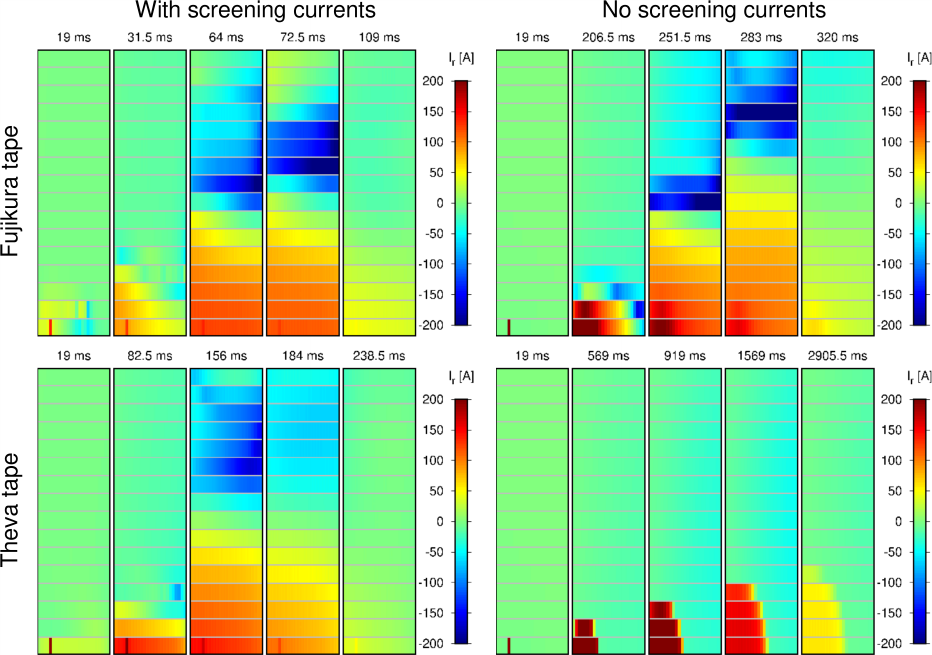}}%
\caption{Radial current between turns appears in the modeled metal-insulation coils during electrothermal quench. \label{f.Irall}}%
\end{figure}

In this section, we compare the inserts made from Theva and Fujikura tapes with $J_c(B,\theta,T)$ in figure \ref{f.JcBT} and we analyze the details of the quench process. The time evolution of the current ($I$), temperature ($T$), quenched zone propagation, the angular current density ($J_\varphi$) and radial current ($I_r$) are in figures \ref{f.Itall}- \ref{f.Irall}. 

For voltage limitation (of 1 V), the current in the REBCO insert decreases with time after the appearance of degraded turns (figure \ref{f.Itall}). This is because several turns in the the cross-section present $I_c$ below the initial input current, and hence there appears an effective resistance. The cause of the decrease of $I_c$ is by both the assumed degradation at time $t_0$ and, more importantly, by the increase in temperature that follows after electrothermal quench. 

Electrothermal quench propagates much faster when screening currents are taken into account for both the Theva and Fujikura tape (figure \ref{f.Tall} and \ref{f.quenchj}). When neglecting screening currents, the insert made of Fujikura tape quenches in the whole cross-section, while for the Theva tape quench remains localized on the bottom-left corner. The causes of this different behavior are the following. First, the initial current is closer to the coil $I_c$ for the Fujikura tape than for the Theva tape. Second, the anisotropy of $J_c$ is very different. While for the Fujikura tape the minimum turn-$I_c$ is at both top and bottom pancakes, for the Theva tape it is at the top-left corner. Then, for the Theva tape the surrounding region of the damaged turns presents turn-$I_c$ well above the minimum $I_c$ in the coil. Another feature is that initially quench propagates from the bottom pancake upwards. However, when around more than half of the pancakes quench, temperature starts to also increase from the top pancake (figure \ref{f.Tall}).

An interesting issue is that the quenched zone for the Theva tape with screening currents shrinks from 189 ms to 238.5 ms (figure \ref{f.quenchj}), using the definition of quenched zone as \R{$|J_\varphi|/J_c\ge 2$}. The reason is that the net current sharply decreases (figure \ref{f.Itall}), and hence \R{$|J_\varphi|/J_c$} can become below 2 due to the decrease of \R{$|J_\varphi|$}. In addition, $T$ stays below $T_c=92$ K in most of the coil, and hence most of the turns remain superconducting. At later times, the quenched zone shrinks further to only a few turns around the damaged sections (not shown). If screening currents are neglected, the quenched zone always remains close to the bottom left corner. For the Fujikura tape, there appears shrinking of the quenched zone (whether screening currents are taken into account or not) although there always remains a substantially large region of quenched cross-section (not shown).

For the angular current density (figure \ref{f.Jphall}), screening currents are evident where quench have not yet occurred. The magnitude of the screening currrents (or maximum $|J_\varphi|$) is higher for the Fujikura tape than Theva because $J_c$ presents a sharp peak for zero perpendicular field to the tape, $B_\perp=0$ or $\theta=90^{\rm o}$ (figure \ref{f.JcBT}). That occurs around the regions with $|J|<J_c$, which are around the center in the $z$ direction of each pancake. On the contrary, for the Theva tape there is no maximum for $\theta=0$, but for $\theta=-30^{\rm o}$. That is why screening currents at the bottom pancakes are of higher magnitude than the top ones. With no screening currents, the changes of angular current can be understood when analyzing the changes in radial current, as detailed below.

For Theva tape with no screening currents, there appears the typical transfer of angular current into radial current in the quenched turns (see figures \ref{f.Jphall} and \ref{f.Irall}). At the non-quenched turns, there appear negative radial currents due to inductive effects, as a response of the decrease in magnetic flux due to the decrease of total current, which causes a decrease of angular current throughout the insert. These negative radial currents increase the radial currents. For the Fujikura insert with no screening currents, there are strong negative radial currents at the pancakes next to the border between the quenched and non-quenched zones. Again, they appear by the decrease of axial magnetic flux, but now due to the decrease of $J_\varphi$ in neighboring regions, instead of a global decrease of $J_\varphi$ due to a decrease in coil current. These radial currents cause high power loss, and hence they are responsible of quench propagation. For screening currents, high negative radial currents still appear next to the quench-non-quenched boundary and high positive radial currents appear in the quenched regions. For all cases, both angular and radial currents are low at the map for the latest time instant in figures \ref{f.Jphall} and \ref{f.Irall} because of the reduction of total current.

\begin{figure}[tbp]
{\includegraphics[trim=79 0 75 10,clip,height=7.5 cm]{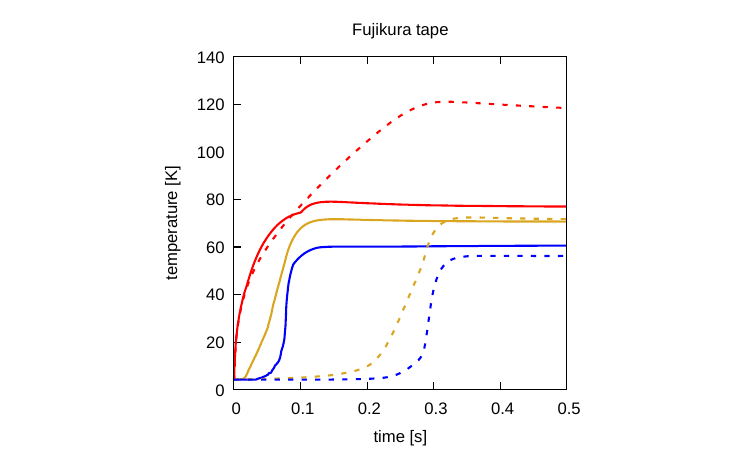}}%
{\includegraphics[trim=90 0 80 10,clip,height=7.5 cm]{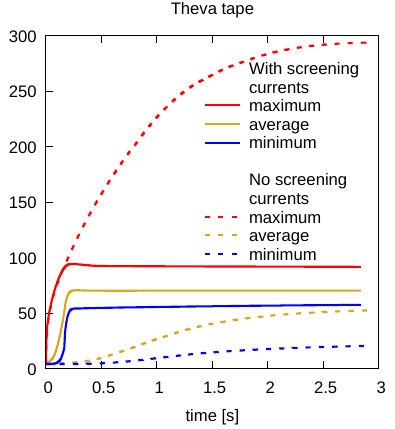}}%
\caption{Comparison of the maximum, average, and minimum temperature in the REBCO inserts when considering either Fujikura or Theva tape. Screening currents reduce the maximum temperature in both cases, but it is higher for the Theva tape. \label{f.Ttall}}%
\end{figure}

Next, we analyze several other global quantities of the inserts.

The time evolution of the maximum, average, and minimum temperature shows that neglecting screening currents predict much higher maximum temperatures (figure \ref{f.Ttall}). For Fujikura tape, the maximum temperature is lower because of faster quench propagation. However, the minimum temperature is lower for the Theva tape. For all cases, the average temperature is well below the critical temperature ($T_c=92$ K), even long after complete quench. Therefore, a substantial portion of the cross-section remains always in superconducting state.

\begin{figure}[tbp]
{\includegraphics[trim=5 33 5 38,clip,width=15.7 cm]{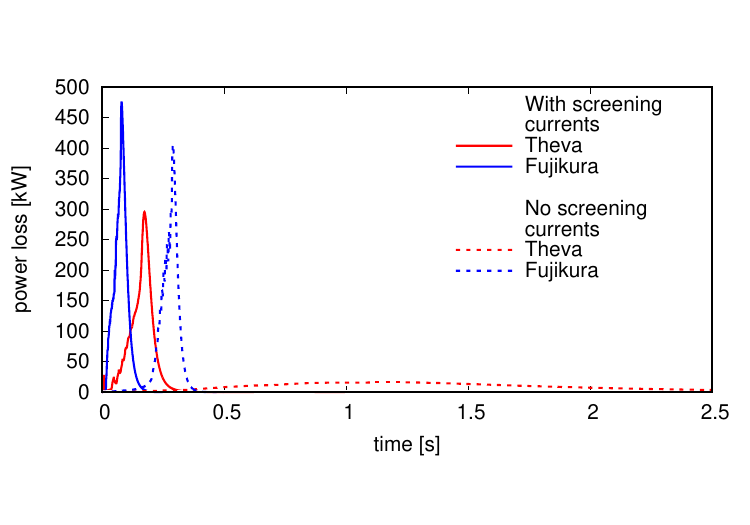}}%
\caption{The stored energy in the REBCO insert dissipates into heat through electromagnetic power loss. The peak power loss is higher with screening currents, specially for the Theva tape. \label{f.Pt}}%
\end{figure}

In modeling, we can identify widespread quench by the generated heat power (figure \ref{f.Pt}). First, the power increases because of the expansion of the quenched region. When the whole coil already quenched, the power decreases because of the decrease in total current. The peak in power highly overcomes the input power, $P_{\rm in}=IV\le 308.8$ W, because of the fast dissipation of the coil magnetic energy. The peak in power roughly corresponds to the instant of complete quench across the coil section, and hence it helps us to compare the quench speed for all cases. Again, quench occurs much earlier when screening currents are taken into account for both Theva and Fujikura tapes.

\begin{figure}[tbp]
{\includegraphics[trim=5 33 5 40,clip,width=15.7 cm]{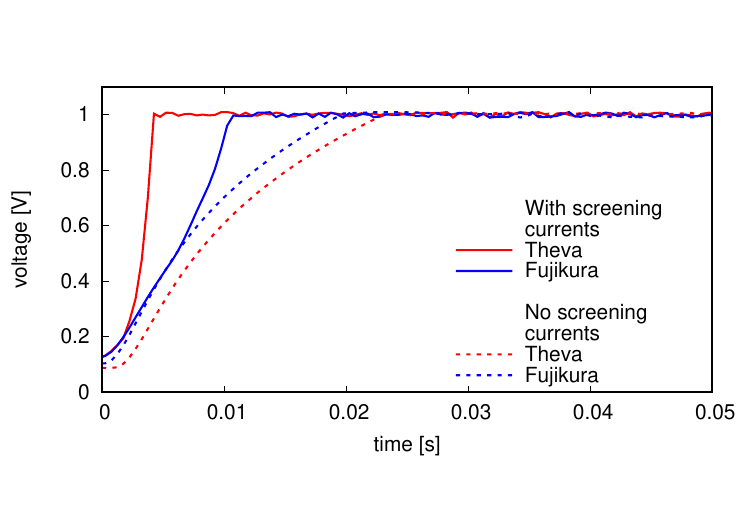}}%
\caption{During quench, the voltage fastly increases until it reaches the limiting value, $V_{\rm lim}=$1 V. \label{f.Vt}}%
\end{figure}

Although voltage limitation acts very soon (between 4 and 23 ms depending on the configuration, figure \ref{f.Vt}), the impact on current is much slower, due to the large coil inductance. In order to experience a sharp decrease in current, a substantial portion of the coil needs to experience $I$ above $I_c$. Quench detection based on voltage, either by 1 V criterion or a portion of it like 0.2 V, occurs shortly after damage (of the order of few ms). However widespread quench appears much later, in the order of hundreds of ms and even around 1 s, the latter for Theva tape with no screening currents. 

\begin{figure}[tbp]
{\includegraphics[trim=2 30 11 35,clip,height=7.5 cm]{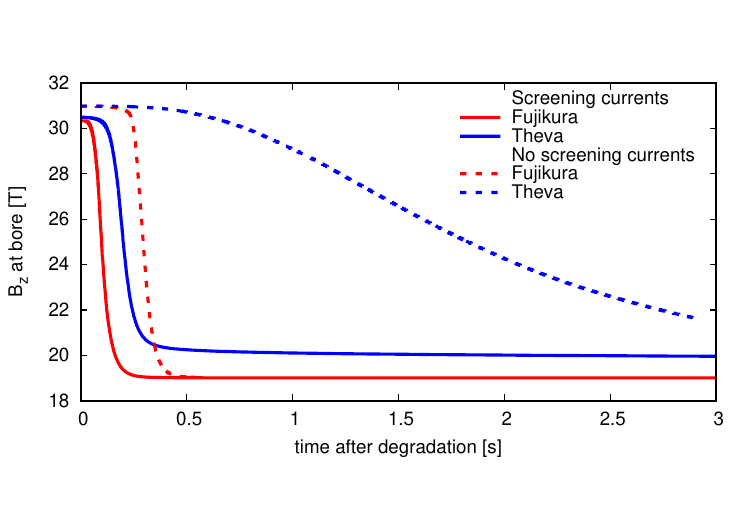}}%
\caption{The generated magnetic field at the bore center decays during quench. Screening currents cause a faster decay, as well as a reduction of the magnetic field at bore due to screening current induced field. \label{f.Bcen}}%
\end{figure}

The magnetic field at the bore center, $B_{z,{\rm cen}}$, decreases with time (figure \ref{f.Bcen}) mainly due to the decrease of net current. However, there is a delay of $B_{z,{\rm cen}}$ compared to $I$ because the decrease in magnetic flux causes negative radial current, which partially retains $J_\varphi$. In this article, $B_{z,{\rm cen}}$ for no screening currents is below 32 T because the operation current (or pre-quench current) is below the rated current (308.8 A compared to 333 A). Screening currents decrease $B_{z,{\rm cen}}$ before the quench event due to Screening Current Induced Field (SCIF). For the Fujikura and Theva inserts SCIF is similar, but for Fujikura it is slightly higher.

\newpage

\newpage

\end{document}